\documentclass[%
reprint,
onecolumn,
amsmath,amssymb,
aps,
floatfix,
notitlepage,
]{revtex4-1}

\usepackage{amssymb,amsmath}
\usepackage{graphicx}
\usepackage{subcaption}
\usepackage{gensymb}
\graphicspath{{fig/}}
\usepackage[colorinlistoftodos]{todonotes}
\usepackage{xcolor}
\usepackage{xfrac}
\usepackage[normalem]{ulem}
\usepackage{tikz}
\usepackage{siunitx}
\usepackage{float}

\usepackage{hyperref}
\hypersetup{
	colorlinks,
	citecolor=blue,
	linkcolor=black,
	urlcolor=black
}

\definecolor{darkolivegreen}{rgb}{0.33, 0.42, 0.18}

\newcommand{\rev}[1]{{#1}}
\newcommand{\revtwo}[1]{{#1}}

\newcommand{\av}[1]{\langle {#1} \rangle}
\newcommand{\vect}[1]{\mathbf{#1}}
\renewcommand{\vec}[1]{\mathbf{#1}}

\usepackage{setspace}
\usepackage[margin=2.5cm]{geometry}

\definecolor{corange}{HTML}{ff7d0c}
\definecolor{cblue}{HTML}{1e75b3}

\begin{document}

\title{Turbulence -- Obstacle Interactions in the Lagrangian Framework: Applications for Stochastic Modeling in Canopy Flows}

\email[]{ronshnapp@gmail.com}

\author{Ron Shnapp}
\altaffiliation{School of Mechanical Engineering, Tel Aviv University, Tel Aviv, Israel}

\author{Yardena~Bohbot-Raviv}
\altaffiliation{Israel Institute for Biological Research, Ness Ziona, Israel}

\author{Alex~Liberzon}
\altaffiliation{School of Mechanical Engineering, Tel Aviv University, Tel Aviv, Israel}

\author{Eyal~Fattal}
\altaffiliation{Israel Institute for Biological Research, Ness Ziona, Israel}





\begin{abstract}
	\revtwo{Lagrangian stochastic models are widely used to predict and analyze turbulent dispersion in complex environments, such as in various terrestrial and marine canopy flows. However, due to a lack of empirical data, it is still not understood how particular features of highly inhomogeneous canopy flows affect the Lagrangian statistics. In this work, we study Lagrangian short time statistics by analyzing empirical Lagrangian trajectories in sub-volumes of space that are small in comparison with the canopy height.
	For the analysis we used 3D Lagrangian trajectories measured in a dense canopy flow model in a wind-tunnel, using an extended version of real-time 3D particle tracking velocimetry (3D-PTV).
	One of our key results is that the random turbulent fluctuations due to the intense dissipation were more dominant than the flow's inhomogeneity in affecting the short-time Lagrangian statistics. This amounts to a so-called quasi-homogeneous regime of Lagrangian statistics at small scales.
    Using the Lagrangian dataset we calculate the Lagrangian autocorrelation function and the second-order Lagrangian structure-function, and extract associated parameters, namely a Lagrangian velocity decorrelation timescale, $T_i$, and the Kolmogorov constant, $C_0$.
    We demonstrate that in the quasi-homogeneous regime, both these functions are well represented using a second-order Lagrangian stochastic model that was designed for homogeneous flows. 
    Furthermore, we show that the spatial variations of the Lagrangian separation of scales, $T_i/\tau_\eta$, and the Kolmogorov constant, $C_0$, cannot be explained by the variation of the Reynolds number, $Re_\lambda$, in space, and that $T_i/\tau_\eta$ was small as compared with homogeneous turbulence predictions at similar $Re_\lambda$.
    We thus hypothesize that these characteristics occurred due to the injection of kinetic energy at small scales due to the so-called ``wake production'' process, and show empirical results supporting our hypothesis.
    These findings shed light on key features of Lagrangian statistics in flows with intense dissipation, and have direct implications for modeling short term dispersion in such complex environments.}
\end{abstract}
\maketitle

\section{Introduction}\label{sec:intro}


Scalar dispersion in the atmospheric surface layer is strongly influenced by the turbulent canopy flows. These flows affect the dispersal of pathogens and the ventilation of urban areas~\cite{Britter2003}, as well as the dispersal of spores, bacteria, and seeds in forests and fields~\cite{Nathan2002}. After years of research, it had become commonly accepted that there is a particular difficulty to model dispersion in canopy flows through Eulerian models, resulting from the failure of gradient-diffusion theory in these flows~\cite{Raupach1981, Raupach1987, Gavze2018}. Consequently, Lagrangian Stochastic Models (LSM) gained popularity as a state-of-the-art modeling approach within the community (e.g. Refs.~\cite{Raupach1989, Flesch1992, Wilson1996, Rotach1996, Baldocchi1997, Leuning2000, Aylor2001, Nathan2002, Arritt2007, Gleicher2014, Duman2016, Fattal2018} among others). LSMs can efficiently predict transport and dispersion in turbulent flows through Monte-Carlo simulations~\cite{Wilson1996}, and are specifically useful in applications to inhomogeneous turbulence (i.e. where the flow statistics vary in space) and to cases with complex distributions of sources.

\revtwo{In the LSM framework, a stochastic Markov random walk provides increments of the states of ``marked'' particles, defined by the particles' position and its time derivatives up to $n$ order $d^n\vec{x}/ dt^n$, (usually $n=1$ or $n=2$ is the order of the model, and boldfaced symbols denote vectors).
There are two fundamental steps in constructing an LSM: a) determine an appropriate stochastic process defining the random walk, and b) to obtain functional expressions relating the Lagrangian operators of the random walk to statistics of the flow field. In other words, one must find Lagrangian equations of motion in terms of a priory known Eulerian velocity statistics.} 

\revtwo{There are significant challenges associated with the two steps above due to the complex nature of the canopy flows, in which various processes occur simultaneously at different scales. In particular, there is a number of topics that demand special attention when constructing LSMs for canopy flows:}
\begin{enumerate}
	\itemsep0em
	\item \revtwo{turbulence generated simultaneously by both the large scale shear and due to flow--obstacle interactions (wakes) by canopy elements, leading to the so-called spectral short-circuiting through wake--production}~\cite{Finnigan2000, Poggi2008};
	\item \revtwo{non uniqueness in the formulations of LSMs for inhomogeneous flows, namely in the solution of Thomson's well-mixed principle}~\cite{Thomson1987,Reynolds1999}; %
	\item \revtwo{inhomogeneity due to the significant difference between the flow above and inside the canopy layer~\cite{Duman2016}, leading to a strong shear layer near the canopy top;}
	\item \revtwo{Uncertainty in the parameterizations of Lagrangian statistics that are needed for LSM construction, for example the Kolmogorof constant or the Lagrangian integral timescale~\cite{Massman1999, Rannik2000, Poggi2006a};}
	\item \revtwo{the prevalence of large coherent structures due to the phenomenon so-called the mixing layer analogy~\cite{Raupach1996}, leading to  non-Gaussian distributions of the Eulerian velocity~\cite{ Shaw1987};}
	\item \revtwo{the ``mechanical diffusion'' as a result of the fluid having to bypass the canopy obstacles~\cite{Nepf1999} (similar to dispersion in porous media flows);}
\end{enumerate}
\revtwo{Notably, the references above could not have been a conclusive list due to the vast body of literature.}
\revtwo{In this work, we focus on small scale motion in wakes of canopy obstacles associated with item 1. Specifically, turbulent kinetic energy is produced in canopies in two ways: production due to mean shear at scales comparable with obstacle height ($H$), and production due to obstacle drag at smaller scales, so-called wake-production. Wake-production is said to short-circuit the turbulent cascade since it extracts energy from large scales and injects it directly at smaller scales~\cite{Finnigan2000}. However, except for the theory by Poggi et al.~\cite{Poggi2008} showing that wake production can affect the Kolmogorov constant $C_0$ (introduced below), how wake production affects Lagrangian statistics in canopy flows is not known. Importantly, direct estimations of Lagrangian flow statistics in canopy flows that could provide crucial data to resolve the items above are lacking.}


Technological and scientific advances of the last two decades enabled the gathering of invaluable empirical data in the Lagrangian framework, both through experiments and direct numerical simulations (e.g. Refs.~\cite{Pope1990, Voth1998, Ott2000, Mordant2002, Biferale2005, Bourgoin2006, Yeung2006a, Ouellette2006, Berg2006, Bec2006, Guala2007, Walpot2007, Arneodo2008, Toschi2009, Liberzon2012, Scatamacchia2012, DiBernardino2017, Stelzenmuller2017, Polanco2018, Shnapp_Liberzon:2018}). Such studies have analyzed Lagrangian dynamics in details, bringing attention in the community to delicate mechanisms underlying the motion of Lagrangian particles in turbulent flows~\cite{Toschi2009}.
The empirical data obtained in such studies can be used to bridge the gap between the fluid mechanics and the stochastic models through more accurate parameterizations and validation of theories. In most cases, due to practical limitations, the focus was put on homogeneous isotropic turbulent flows (HIT); however, in the recent years, studies with embedded inhomogeneity and anisotropy are gaining more attention (e.g. Refs.~\cite{Celani2005, Pitton2012, DiBernardino2017, Stelzenmuller2017, Polanco2018,Shnapp2019}).

\revtwo{In this study, we use measurements of flow tracers' trajectories in a heterogeneous canopy modeled in a wind tunnel,~\cite{Shnapp2019, Bohbot-Raviv2017}, to better understand the effect of flow--obstacle interactions on short time Lagrangian statistics (item 1). We focus our analysis on small scales by exploring Lagrangian statistics inside small sub-volumes of space, namely smaller than $\frac{\av{u}}{\partial \av{u} / \partial z}$ where $\av{u}$ is the streamwise mean velocity. Our analysis reveals that the short time Lagrangian statistics were affected more strongly by random turbulent fluctuations than by the flow inhomogeneity due to intense dissipation; this amounts to quasi-homogeneous regime of Lagrangian dynamics at short-time and small-scale. Accordingly, the Lagrangian autocorrelation function and the second-order structure-function could be well represented by a second order LSM designed for homogeneous flows. Furthermore, our empirical study in the quasi-homogeneous regime suggests that Lagrangian separation of scales and the Kolmogorov constant (definitions below) are affected by wake-production, in agreement with Poggi et al.~\cite{Poggi2008}. The results of our study are relevant for flows with intense dissipation and spectral short-circuiting effects. Importantly, we demonstrate how measurements performed directly in the Lagrangian framework can provide crucial insight into Lagrangian dynamics in canopies, and thus that they have potential for improving dispersion models' accuracy in canopy flows.}

\revtwo{The rest of this paper is organized as follows. In Section~\ref{sec:methods} we present three LSMs that will be used in our analysis, and the details of our experiment and our analysis. In Section~\ref{sec:Quasi-homogeneity}, we reveal a quasi-homogeneous regime of Lagrangian statistics in the small scales. In Section~\ref{sec:res} we present direct estimations of the Lagrangian velocity autocorrelation, the structure-functions and the associated parameters ($T_i$ and $C_0$), demonstrating their spatial distributions. In Section~\ref{sec:rapid_decorelation}, we compare our empirical results with predictions for homogeneous flows and hypothesize on the role of wake production. Lastly, we summarize and present our conclusions in Section~\ref{sec:conclusions}.}


\section{Methods}\label{sec:methods}

\subsection{Definitions}\label{sec:three_LSMs}

Let us first define three different LSMs that will be used in the consequent analysis. The simplest first order LSM ($n=1$) assumes that the turbulence has spatially homogeneous statistics and that it is characterized by Gaussian velocity PDFs at each point (hereafter called Gaussian turbulence). This model is essentially the Ornstein-Uhlenbeck (OU) process~\cite{Gardiner1997, Wilson1996}:
\begin{equation}
dv'_i = \alpha_i(\mathbf{x}, \mathbf{v},t) \, dt  +  \beta_{ij}(\mathbf{x}, \mathbf{v}, t) \, d \xi_j \quad ; \quad dx_i = \, \left[ v'_i + \av{u_i} \right] \, d\tau
\label{eq:langevin}
\end{equation}
\begin{equation}
\alpha_i = - \frac{v'_i}{T_{L,i}}    \quad ; \quad  \beta_{ij}  = \delta_{ij}\,\left( \frac{2 \sigma_i^2}{T_{L,i}} \right)^{1/2}.
\label{eq:langevin_homogeneous}
\end{equation}
Here $v_i = \av{u_i} + v'_i$ is the $i^\mathrm{th}$ component of the velocity of a Lagrangian particle where $\av{u_i}$ is the mean velocity field and $v_i'$ a velocity fluctuation, $d \xi_i$ are increment components of a Wiener process, $\delta_{ij}$ is the Kronecker delta function, and $\sigma_i$ is the standard deviation of $v_i'$. The Lagrangian velocity autocorrelation function, $\rho_{ij}$, and the Lagrangian integral timescale, $T_{L,i}$, are defined as:
\begin{equation}
\rho_{ij}(\tau) = \frac{\av{v'_i(0) \, v'_j(\tau)}}{\av{v_i^{'2}(0)}^{1/2} \, \, \av{v_j^{'2}(\tau)}^{1/2}}
\quad ; \quad
T_{L,i} = \int_0^{\infty}\rho_{ii}(\tau) d\tau
\label{eq:autocorrelation}
\end{equation}
\noindent where angular brackets $\av{ \cdot }$ denote an average, the velocities $v_i(0) \, v_j(\tau)$ are taken along an individual Lagrangian trajectory, and since we are considering a homogeneous and stationary flow, $\rho_{ij}(\tau)$ is a function of $\tau$ only. 
Doob \cite{Doob1942} proved that the OU process, Eq.~\eqref{eq:langevin}, is essentially the only process defined with the properties: stationary, Gaussian and Markovian, with an exponential autocorrelation function:
\begin{equation}
\rho_{ij}(\tau) = \delta_{ij} \exp\left(-\frac{\tau}{T_{L,i}}\right) .
\label{eq:thomson_model}
\end{equation}
Furthermore, based on the Obukhov conjecture \cite{Obukhov1959}, consistency with the Kolmogorov inertial range scaling of Lagrangian velocity increments~\cite{Monin1972} requires that~\cite{Wilson1996} --
\begin{equation}
\begin{split}
\frac{2\,\sigma_i^2}{T_{L,i}}  = \frac{\, D_{ii}(\tau)}{\tau}  = C_0 \, \epsilon \quad ; \quad D_{ii}(\tau)  = \av{ \left[ v'_i(\tau) - v'_i(0) \right]^2}
\end{split}
\label{eq:b}
\end{equation}  
where $D_{ii}(\tau)$ is termed the Lagrangian second order structure function, $C_0$ is the so-called Kolmogorov constant, and $\epsilon$ is the mean rate of turbulent kinetic energy dissipation. Borgas and Sawford~\cite{Borgas1994} showed that for HIT these choices of $\alpha_i, \, \beta_{i}$ are unique solutions of Thomson's well-mixed condition~\cite{Thomson1987} and thus it is an exact solution of the Fokker-Plank equation.

\revtwo{In realistic situations the flows' statistics vary in space, namely flows are inhomogeneous, and this fact is not treated by the simplest model above, eq.~\eqref{eq:langevin_homogeneous}. To overcome this issue, Thomson~\cite{Thomson1987} solved the well-mixed condition and provided a LSM for such flows assuming that the turbulence is Gaussian and consistent with Kolmogorov similarity for $D_{ij}$. This Markovian LSM for inhomogeneous flows still uses eq.~\eqref{eq:langevin}, but the coefficients change according to:}
\begin{equation}
\begin{split}
&\alpha_i = -\frac{1}{2} C_0 \epsilon \, R^{-1}_{ij} \,v'_j + \frac{\phi_i}{g} \quad ; \quad b_{ij} = \sqrt{C_0 \, \epsilon} \delta_{ij}\\[.5em]
\frac{\phi_i}{g} = 
&\frac{1}{2}\,\frac{\partial R_{il}}{\partial x_l} +
\frac{1}{2}R^{-1}_{lj} \, \av{u_k} \, \frac{\partial R_{il}}{\partial x_k} \, v'_j + 
\frac{1}{2}R^{-1}_{lj} \, \frac{\partial R_{il}}{\partial x_k} \, v'_j v'_k \,\, .
\end{split}
\label{eq:phi}
\end{equation}
\revtwo{Following Wilson et al.~\cite{Wilson2009}, this equation can be written more compactly as:}
\begin{equation}
\frac{\phi_i}{g} = T^0_i + T^1_{ij} \, v'_j + T^2_{ijk} \, v'_j\, v'_k
\label{eq:inhom_a} 
\end{equation}
\revtwo{where the $T^m$ are coefficients multiplying the fluctuating velocity to power $m$. As written in the introduction (item 2), this model is not a unique solution of the well-mixed condition under the above assumptions. Nevertheless, it is commonly used for modeling dispersion in canopy flows (e.g., Refs~\cite{Baldocchi1997, Poggi2006a, Wilson2009, Duman2014}) and we shall utilize it here as well.}

So far we have only considered first order LSMs ($n=1$) that are regarded accurate in flows with very high Reynolds numbers. In such cases the particle accelerations have very short correlation times relative to the integral timescales~\cite{Monin1972, Pope1990,Voth1998}. However, if this separation of timescales reduces, for example due to finite Reynolds number effects, the first order models become less accurate in modeling the dispersion. The influence of finite separation of scales was addressed by Sawford~\cite{Sawford1991} in the introduction of a second order model ($n=2$) for a homogeneous turbulence case: 
\begin{equation}
\begin{split}
T_L da_i + ( 1 + R^{1/2})  \, a_i(t) \, dt + \frac{R^{1/2}}{T_L} \int_0^t & a_i(\tau) \, d\tau \, dt  =  \sqrt{\frac{2 \sigma}{T_L} R^{1/2} \left( 1 + R^{-1/2} \right)} \, d\xi_i  \\[.15em]
dv_i = a_i \, d\tau \,\,  ; \,\, dx_i = (v_i& + \av{u_i}) \, d\tau \,\, ; \,\, R = \frac{16 a_0^2}{C_0^4} \left(\frac{\tau_e}{\tau_\eta}\right)^2 
\end{split}
\label{eq:sawford91}
\end{equation}
where $a_0$ is a Kolmogorov constant for the variance of acceleration, $a_i$ is a component of the particle's acceleration vector, $\tau_e$ is an Eulerian integral timescale, and $\tau_\eta=(\nu/\epsilon)^{1/2}$ is the Kolmogorov timescale. The model Eq.~\eqref{eq:sawford91} has the following velocity autocorrelation function~\cite{Sawford1991}:
\begin{equation}
\rho_{ij}(\tau) = \delta_{ij} \, \frac{ T_{1,i} \exp(-\tau / T_{2,i}) - T_{2,i} \exp(-\tau / T_{1,i})}{T_{1,i} - T_{2,i}}  \quad ; \quad T_{L,i} = T_{1,i} + T_{2,i} \, . 
\label{eq:sawford_model}
\end{equation}
This second order model was extended by Du et al.~\cite{Du1995} to the decaying turbulence case and 
by Reynolds~\cite{Reynolds1999} to include vertical inhomogeneity of the turbulent flow statistics.

Above we have considered three models: Eq.~\eqref{eq:langevin} with Eq.~\eqref{eq:langevin_homogeneous} which is a first order model for infinite Reynolds number homogeneous turbulence, Eq.~\eqref{eq:langevin} with Eq.~\eqref{eq:inhom_a} which is also a first order model but for inhomogeneous flows, and lastly, Eq.~\eqref{eq:sawford91}, which is a second order model, incorporating finite Reynolds number effects in homogeneous flows. These models will be used in the analysis of our results.

\subsection{Experimental Method}

In our investigation, we used the results from a wind-tunnel experiment. Detailed descriptions of the experimental apparatus, measurements, and post processing, were given in Refs.~\cite{Shnapp2019, Bohbot-Raviv2017}, yet for completeness, we present a brief overview here as well. We used the environmental wind-tunnel laboratory at IIBR, featuring a 14 meters long open wind-tunnel with a $ 2 \times 2\, \si{\meter\squared}$ cross sectional area, that is compatible for conducting experiments mimicking turbulent flows in the atmospheric surface layer. The canopy flow was modeled by placing flat rectangular plates on the bottom floor of the wind tunnel. Our mixed height canopy layer was constructed of two types of plates with a height of either $H$ or $\frac{1}{2}H$, and a width of $\frac{1}{2}H$, where $H=100\, \si{\milli \meter}$. The two types of plates were positioned in consecutive rows and at a staggered orientation, see Fig~\ref{fig:locs}(a) and (b). The entire upstream part of the test section was fitted with roughness elements. The canopy frontal area index, defined as the plate frontal area divided by the lot area, $\Lambda_f = A_f /A_T=\sfrac{9}{16}$, ($A_f$ being is the element frontal area, and $A_T$ the lot area of the canopy). These densities categorizes our canopy as a  moderately dense and deep canopy.

We gathered data at two levels of the free stream velocities, corresponding to Reynolds numbers $Re_\infty \equiv U_\infty H/\nu = 16 \times 10^3 $ and $26 \times 10^3$; with $U_\infty = 2.5$ and $4 \, \si{\meter \per \second}$ being the free stream mean velocity measured with a sonic anemometer at the center of the wind-tunnel cross section, and $\nu$ the kinematic viscosity of the air. In what follows we adopt the frame of reference commonly used in the canopy flow literature -- $x$ is streamwise aligned longitudinally within the wind-tunnel, $y$ is in the horizontal cross-stream direction, and the positive  $z$ axis is directed vertically away from the bottom wall at which $z=0$.

In the experiment, we tracked fluid tracers using a real-time extension of the 3D-PTV method~\cite{Shnapp2019}. The flow was seeded with hollow glass spheres of diameter 5 micrometer on average, with the Stokes number $St= \tau_p/\tau_\eta \approx 0.05$. The tracers were illuminated with a 10W, 532nm continuous wave laser; 3D positions were inferred from 4 Mega pixel images at a resolution of 50 $\si{\micro \meter}$ per pixel and rates of 500 $\si{\hertz}$ within the canopy layer $z\leq H$, and 1000 $\si{\hertz}$ above the elements $z > H$. We implemented camera calibration, stereo matching and tracking~\cite{Dracos1996}, to reconstruct the tracer particle's trajectories by using the OpenPTV~\cite{openptv} open source software, integrated to operate with the real-time image analysis extension. The trajectory data analysis was performed by employing our open source Flowtracks package~\cite{Meller2016}.

\subsection{The Sub-Volume Approach}\label{sec:sub_volume_approach}

\begin{figure}[t]
	\centering
	\includegraphics[width=13cm]{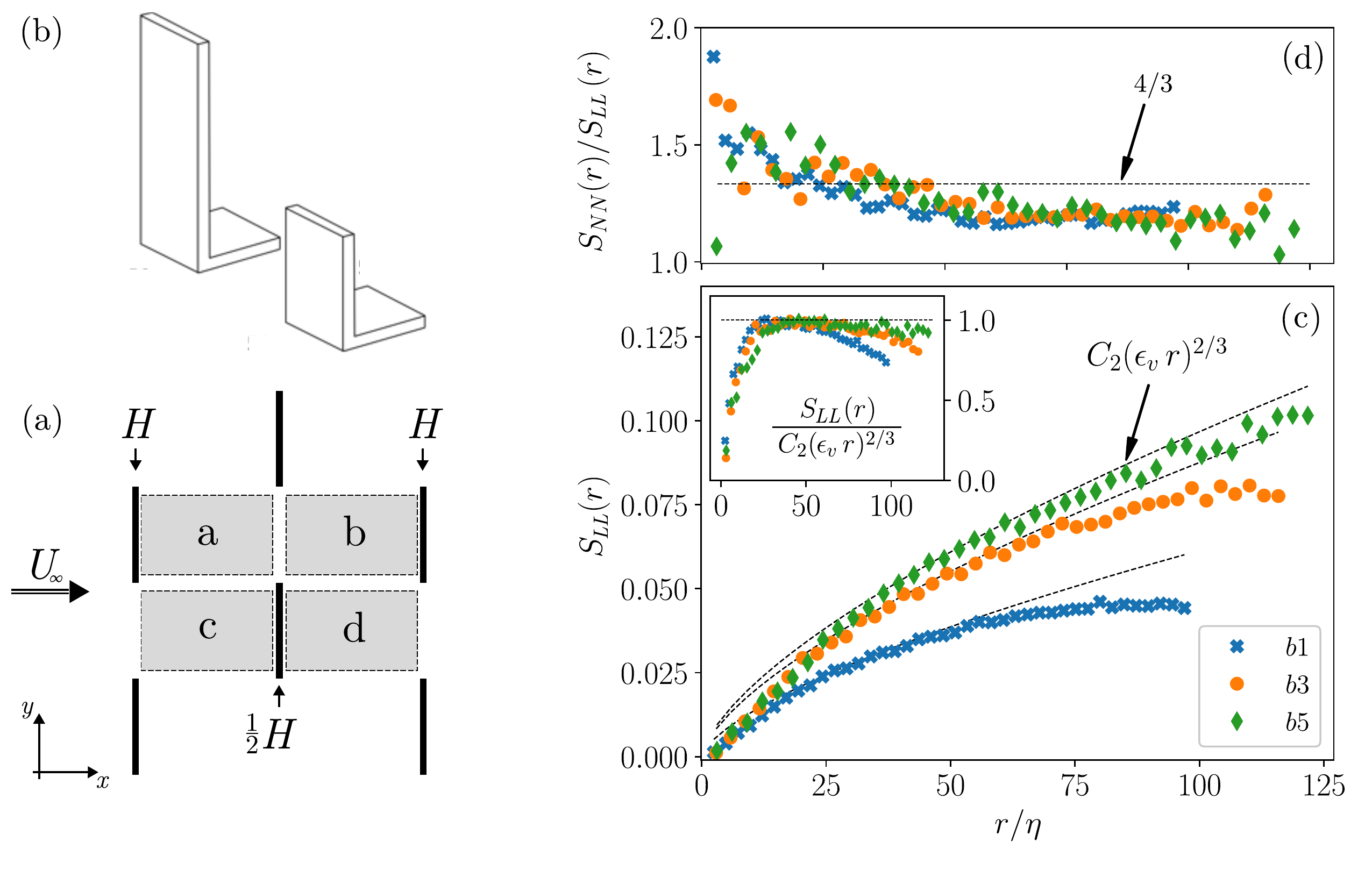}
	\caption{\rev{(a) A top view sketch of the canopy repeating unit cell. Grey shaded regions show the positions of the four sub-volumes. Black thin rectangles represent the canopy roughness elements. (b) An isometric sketch of the short and tall roughness obstacles used. (c) Main panel shows the second order Eulerian longitudinal structure function in three sub-volumes, and dashed lines show the isotropic model, Eq.~\eqref{eq:S2}, with the estimated values $\epsilon_{v}$. The inset shows the compensated structure functions for the same cases. (d) The ratio between the transverse and the longitudinal Eulerian structure functions, compared with the K41 isotropic value $4/3$. \label{fig:locs}}}
\end{figure}

We analyzed trajectories in 20 sub-volumes centered at different locations. The sub-volumes are rectangular cuboids found at 4 horizontal locations, (see diagram in Fig.~\ref{fig:locs}(a)), and at 5 different heights above the wind-tunnel bottom wall. The 4 horizontal positions of the sub-volumes are labeled alphabetically $a$, $b$, $c$ and $d$; $a$ is found immediately downstream of a tall element, $b$ is upstream to the next tall element, and $c$ and $d$ are parallel to the former elements and positioned around a short element. At each horizontal position, $a-d$, we used 5 vertical slabs of thickness $\delta z = 0.2 H$ which defines a total of 20 sub-volumes; the vertical slab position is labeled numerically 1-5 which correspond to the heights 0.5-0.7, 0.7-0.9, 0.9-1.1, 1.1-1.3 and 1.3-1.5$H$, respectively. Thus, for example, the sub-volume $b2$ is located upstream of a tall element at height $0.7H<z\le0.9H$. An animation of a sub-sample from our data set can be seen online through the link~\cite{Shnapp2018a}.

Lagrangian statistics are commonly represented based on a common point of origin $(x_0,t_0)$ and as a function of time~\cite{Monin1972}. Thus the Lagrangian mean of an arbitrary function $\av{A(x_0,t_0,t)}$, is normally defined over flow ensembles. In this work, we assume that the flow is stationary (see~\cite{Shnapp2019}) and ergodic, and thus we replace the above with an average over $t_0$, presented as a function of time lag $\tau = t - t_0$, where $t_0$ is the time a particle was first spotted in our measurement volume. In addition, we present Lagrangian statistics for particles with $\vec{x}_0$ in each of the sub-volumes, namely statistics are sub-volume averaged. 
Therefore, the ensemble average over sub-volume $\mathbb{V}$, $\langle A \rangle_{v} $ of any quantity $A$, is calculated as:
\begin{equation}
\langle A(\vec{x},\tau)\rangle_{v}  \equiv \frac{1}{N}\sum\limits_{i=0}^N A_i(\vec{x}_{v},\tau)
\label{eq:mean}
\end{equation} 
such that $\vec{x}_{v}$ are positions of Lagrangian tracers inside the sub-volume $\vec{x}_v \in \mathbb{V}$, and $N$ is the number of Lagrangian trajectories in the sub-volume over the measurement period of time. 

\revtwo{It is notable that the sub-volumes we used are small. In particular, $\delta z < \frac{\av{u_x}}{\partial \av{u_x} / \partial z}$, and we observed that the flow's statistics had only minor variations within each sub-volumes. This frames the focus of our work on small scale Lagranigan statistics.} 

\subsection{Estimating the Mean Rate of Dissipation}\label{sec:dissipation}

The Eulerian second order structure function is defined as the second moment of spatial velocity differences $\delta \vec{u}'_r \equiv \vec{u}'(\vec{x}) - \vec{u}'(\vec{x}+\vec{r})$~\cite{Monin1972}. \rev{Assuming local isotropy and homogeneity, the Kolmogorov universal similarity theory~\cite{Kolmogorov1941} predicts that the longitudinal component of the second order structure function, $S_{LL}$, and its transverse component, $S_{NN}$, (i.e. the components aligned with $\vec{r}$ or normal to $\vec{r}$ respectively) admit to the following scaling law in the inertial range of scales:
	\begin{equation}
	\begin{split}
	S_{LL}(\vect{x},r) &= \left\langle \left(\delta \vect{u}'_r \cdot \frac{\vect{r}}{r}\right)^2 \right\rangle = C_2 (\epsilon r)^{2/3}  \\[.15em]
	S_{NN}(\vect{x},r) = \frac{1}{2}& \Big( \av{\delta \vec{v}^2(\vec{r})} - S_{LL}(r) \Big) = \frac{4}{3} C_2 (\epsilon r)^{2/3}
	\end{split}
	\label{eq:S2}
	\end{equation}}
where in the homogeneous turbulence case the $\vec{x}$ dependence drops, $r=|\vec{r}|$, and $C_2 \approx 2.1$ is a supposedly universal constant~\cite{Sreenivasan1995}.

Following Refs.~\cite{Ott2000, Shnapp_Liberzon:2018}, we implemented Eq.~\eqref{eq:S2} to estimate $S_{LL}$ in each sub-volume using the Lagrangian dataset and averaging velocity differences over spherical shells. This gives a structure function that is isotropic by construction, namely it depends only on the distance $r$. For example, in the main panel of Fig.~\ref{fig:locs}(c) we present our estimations of $S_{LL}(r)$ in the sub-volumes b1, b3 and b5 for the $Re_\infty=16\times 10^3$ case. Note that our estimation of $S_{LL}$ does not use the Taylor's hypothesis.


Using $S_{LL}$ we estimated an empirical mean rate of dissipation in each sub-volume.
Since $S_{LL}$ is quadratic with $r$ in the dissipation range and should change very slowly at large scales above the integral scale $\mathcal{L}$~\cite{Monin1972}, the compensated structure functions $S_{LL}(r) / C_2 (\epsilon \, r)^{2/3}$ should peak at an intermediate range $\eta<r<\mathcal{L}$, and this is shown in the inset of Fig.~\ref{fig:locs}(c). Thus, we defined a sub-volume averaged dissipation rate as
\begin{equation}
\epsilon_{v} \equiv \max_r \left[ \frac{S_{LL}(r)}{ C_2 \, r^{2/3} } \right]
\label{eq:epsilon}
\end{equation} 
where $S_{LL}$ is estimated with samples $\vec{x} \in \mathbb{V}$.
Using Eq.~\eqref{eq:epsilon} instead of a least square fitting has the advantage of not having to specify a range of $r$ where an inertial range scaling supposedly exists; on the other hand, it may overestimate the value that would have been obtained in a fitting process, but this uncertainty is low (arguably much lower than the uncertainty in the value of $C_2$). Also, due to anisotropy and inhomogeneity, the applicability of Eq.~\eqref{eq:S2} to canopy flow turbulence is not straightforward,  as observed for example in Refs.~\cite{Chamecki2004, Poggi2006a, Poggi2010}. Nevertheless, in Fig.~\ref{fig:locs}(c) we compare our estimations  of $S_{LL}(r)$ with Eq.~\eqref{eq:S2} using $\epsilon_v$, and for the three cases we observe a distinct range of $r$, in which an agreement between the theory and the empirical data exists. 
In addition to that, the transverse components, $S_{NN}$ were similarly estimated using Eq.~\eqref{eq:S2} and the Lagrangian dataset. The ratios $S_{NN}/S_{LL}$ were calculated and an example is shown in Fig.~\ref{fig:locs}(d) as a function of $r$ for the three sub-volumes b1, b3, and b5. The K41 isotropic value of $S_{NN} = \frac{4}{3} S_{LL}$~\cite{Kolmogorov1941} is plotted for comparison as well. The figure shows that for all three sub-volumes, the ratio $S_{NN}/S_{LL}$ decreases with $r$ while crossing the 4/3 value at a limited range of $r/\eta$. These observations of agreement with K41 similarity in a certain limited range of $r$ supports the use of $\epsilon_v$ as a parameterization in our experiment. Therefore, sub-volume averaged dissipation scales were calculated as $\eta \equiv \left( \nu^3 / \epsilon_{v} \right)^{1/4}$ and $\tau_\eta \equiv \left( \nu / \epsilon_{v} \right)^{1/2}$ being the length and time scales respectively, in analogy to the usual case~\cite{Pope2000}.

With the estimations of $\epsilon_{v}$, a Taylor microscale Reynolds number is defined as --
\begin{equation}
Re_{\lambda,v} \equiv v'^2 \left( \frac{15}{\epsilon_{v} \, \nu} \right)^{1/2}
\label{eq:Re_lambda}
\end{equation}
where $v'^2=\av{\sum_i (v_i - \av{u_i})^2}$ is the RMS of particle velocity relative to the sub-volume averaged velocity. Estimations of $\epsilon_{v}$ and $Re_{\lambda,v}$ were repeated for all sub-volumes and the values are tabulated in the Appendix. The Reynolds numbers $Re_{\lambda,v}$ varied with $x$ in the range 350-850 due to the flow's inhomogeneity, and had negligible dependence on $Re_\infty$.
The obtained values were used for parameterization of the results presented below and the subscript ``$v$'' is omitted to facilitate readability.


\section{Quasi-homogeneous Lagrangian Dynamics in the small scales}\label{sec:Quasi-homogeneity}

\revtwo{In this section, we determine the dominant dynamical factors in the context of our measurements by quantifying and comparing contributions from the different terms in the LSM eq.~\eqref{eq:inhom_a}. 
Identifying the most relevant processes in the context our Lagrangian measurements is crucial since it will present our results in the proper context. In addition to that, it will dictate the tools that we use in the consecutive analysis.
}

\revtwo{For the purpose of this analysis, we use the Thomson formulation~\cite{Thomson1987} LSM for inhomogeneous turbulence, eq.~\eqref{eq:langevin} and eq.~\eqref{eq:inhom_a}. Although the LSM is an equation of motion in the Lagrangian framework, its input is a set of Eulerian variables, so to use it we must cast our Lagrangian measurements onto an Eulerian coordinate system. Specifically, we require field representations of the mean velocity, $\av{u_i}$, the turbulent stress tensor, $R_{ij}$, and the Lagrangian structure function parameterization, $C_0\, \epsilon$, all of which are available from our empirical dataset. Thus, we estimate the velocity statistics by using the sub-volume averages: $\av{u_i} = \av{v_i}$ and $R_{ij}=\av{v'_i \, v'_j}$. For $C_0 \, \epsilon$, we rely on the fact that through our Lagrangian measurements we can calculate $D_{ij}(\tau)$ directly from the definition, Eq.~\eqref{eq:b}. This is unlike Eulerian measurements that must rely on models or on previous measurements in similar flows. Thus, we calculate the structure function parameterization by $C_0 \, \epsilon = \max_\tau [\frac{D_{ii}}{\tau}]$; this issue is discussed in detail below. Following that, we obtain continuous and differentiable field estimations by using inverse-distance weighted interpolations of the sub-volume averaged data (for details, see Appendix D). Through this interpolation scheme, we obtained estimations of the Eulerian fields in our volume of measurement, that allow to calculate the models' coefficients.} 

\begin{figure}
	\centering
	\includegraphics[width=11cm]{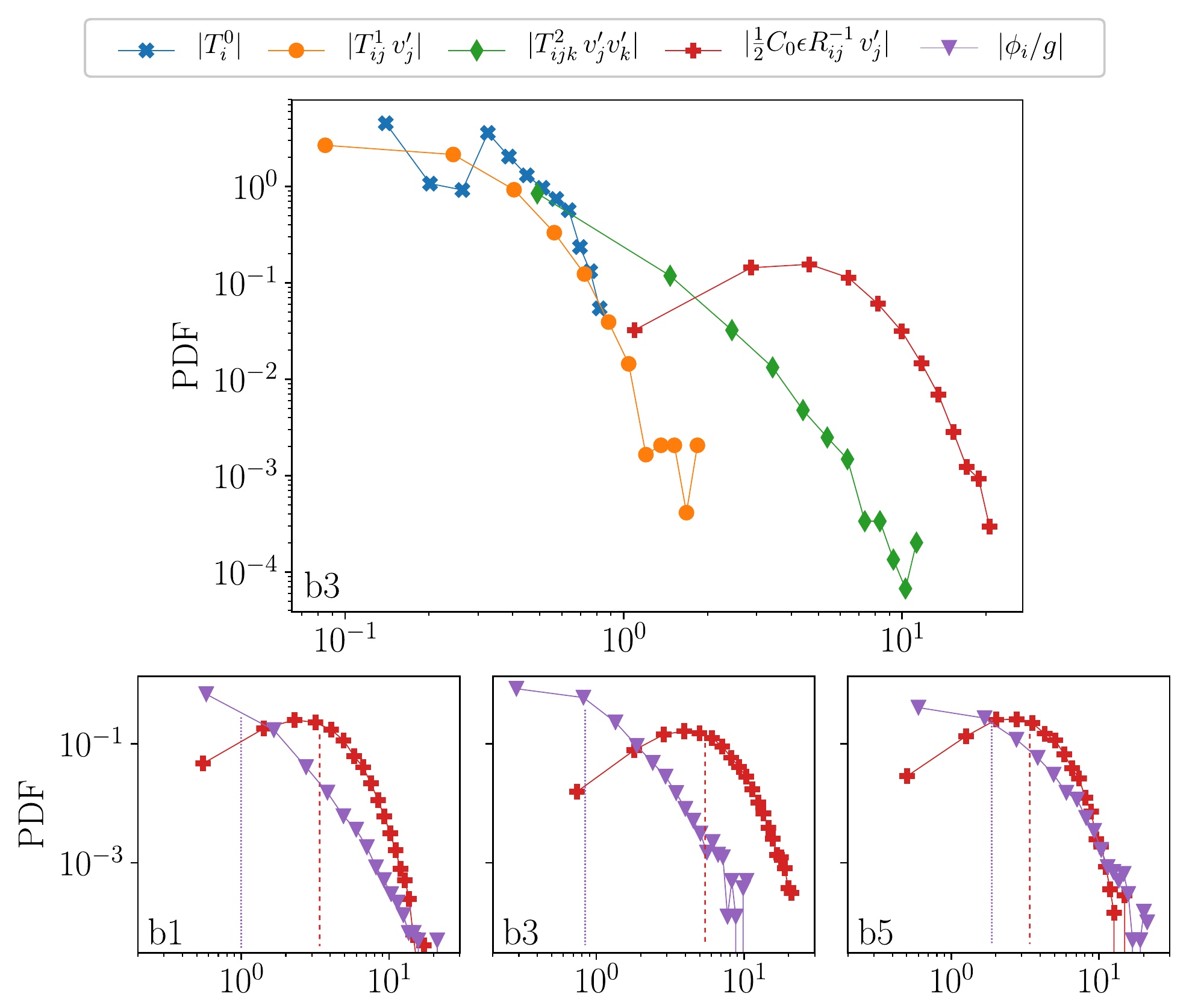}
	\caption{\revtwo{Probability distribution functions (PDFs), where the horizontal axis stands for magnitude of various inhomogeneous LSM terms shown in the legend, sampled along the Lagrangian trajectories from our dataset. Data corresponding to all the model's terms in sub-volume b3, representing $0.9H < z \leq 1.1H$, is shown in the top panel. PDFs for the sum of flow inhomogeneity contributions ($\phi_i/g$) in three sub-volumes, representing regions inside (b1), at the top of (b3), and above (b5), the canopy layer, is shown in the bottom panel. Vertical dotted and dashed lines mark the mean values of the PDFs. For all cases $Re_\infty=1.6\times10^4$.}
		\label{fig:model_coefficients_pdf}}
\end{figure}

\revtwo{We compare the magnitudes of the various LSM terms in Eq.~\eqref{eq:inhom_a} along the trajectories from our experiment using the above interpolation schemes . In particular, to facilitate the analysis, we employ the Wilson et al.~\cite{Wilson2009} formulation for the models coefficients:
$$T_i^0 = \frac{1}{2}\frac{\partial R_{ij}}{\partial x_j} \quad ; \quad  
T_{ijk}^2 = \frac{1}{2} R_{lj}^{-1} \frac{\partial R_{il}}{\partial x_k} \quad ; \quad 
T_{ij}^1 = T_{ijk}^2 \, \av{u_k} $$
In the top panel of Fig.~\ref{fig:model_coefficients_pdf} we show probability distributions for the magnitude of the various LSM terms along the trajectories in our empirical dataset in sub-volume b3, namely at height $0.9 < \frac{z}{H} \leq 1.1$. The curves essentially show the magnitudes of the forces that acted on our measured particles, decomposed according to the LSM. The $T_i^0$  term accounts for inter sub-volume variations of the Reynolds stress tensor, and it has the narrowest distribution of all the terms. The $T^1_{ij}$ and $T^2_{ijk}$ terms are characterized by much wider PDFs since they include both spatial variations of $R_{ij}$ and the randomness of temporal velocity fluctuations. The figure also shows that the term accounting for relaxation of turbulent fluctuations due to dissipation, $\frac{1}{2} C_0 \epsilon R_{ij}^{-1}$, is the largest of all the terms. This implies that velocity changes due to effects of random turbulent fluctuations were generally more dominant than the effects of all the other force components. Notably, the $T^m$ terms arise due to inhomogeneity of the flow, while the $\frac{1}{2} C_0 \epsilon R_{ij}^{-1}$ term accounts for turbulent fluctuations that exsist in both homogeneous and inhomogeneous flows (Eq.~\eqref{eq:langevin_homogeneous}). Thus, in the bottom panel of Fig.~\ref{fig:model_coefficients_pdf} we compare PDFs of $|\phi/g| = |T_{i}^0 + T_{ij}^1 \, v'_j + T_{ijk}^2 \, v'_j v'_k|$ that represents the magnitued of the combined effect of flow inhomogeneity, with that of the dissipation term. The data is shown for three sub-volumes representing the regions inside, right at top of, and above, the canopy layer (b1, b3, and b5 respectively). The mean of each PDF is also marked with a vertical line. The figure shows that for all the regions tested the relaxation due to random fluctuations were more dominant than the effect of flow inhomogeneity. In particular, the ratio between the mean value of the terms was $\av{ \frac{|C_0 \epsilon R_{ij}^{-1} v'_j   |}{2 |\phi_i / g|}  } \approx 3.4$ in sub-volume b1, it increased to 6.4 in sub-volume b3, and it was the smallest, roughly 1.8, above the canopy in sub-volume b5, suggesting that above the canopy the dominance of the dissipation over the effects of inhomogeneity reduced.}

\revtwo{Throughout the entire volume of measurement, we observed that inside and close to the top of the canopy layer $|\phi_i/g|$ was generally smaller than the homogeneous term, $|\frac{1}{2}C_0 \epsilon R_{ij}^{-1} v'_j|$. This means that random forces due turbulent dissipation were stronger than forces due to flow inhomogeneity. In regions farther above the canopy layer (i.e. sub-volumes 4 and 5 with $z>1.1H$), the difference between the two terms reduced, although the turbulent fluctuations through $|\frac{1}{2}C_0 \epsilon R_{ij}^{-1} v'_j|$ were still significantly more dominant. In comparing the variation of the terms with $z$ we observed that $|\frac{1}{2}C_0 \epsilon R_{ij}^{-1} v'_j|$ typically increased with $z$ inside the canopy, peaked at the top of the canopy, and than reduced with $z$ above the canopy; on the other hand, both $|T^1_{ij}\,v'_j|$ and $|T^2_{ij}\,v'_j|$ became increasingly stronger with the height. Thus, as the distance from the wall increases the effects of dissipation became weaker and those of inhomogeneity became stronger. This suggests that the reason for this effect is the direct interaction of the flow with canopy obstacles, that its effects are weaker above the canopy. The flow-obstacle interaction is known to cause an increased energy production and dissipation due canopy drag and wake production~\cite{Finnigan2000}, and seems to dominate the effects of spatial variations of flow statistics in the canopy obstacles' wake region.}

\revtwo{As described above, our analysis of Lagrangian statistics was performed in sub-volumes of space that were much smaller than the integral scale of the flow (i.e. of the order $H$), and indeed the flow statistics did not change appreciably inside each sub-volume. Furthermore, the results presented in this section show unequivocally that the flow inhomogeneity did not have a dominant effect on the Lagrangian dynamics. This leads to the conclusion that there is a quasi-homogeneous regime of Lagrangian statistics at small times, since the contributions from flow inhomogeneity are negligible at small scales as compared to turbulent dissipation. It is important to stress that the canopy flow is inherently inhomogeneous since, for example, statistics change from one sub-volume to another. Nevertheless, inhomogeneity effects on short time Lagrangian statistics, i.e. in scales $\ll \frac{\av{u_x}}{\partial \av{u_x} / \partial z}$, were sub-dominant as compared to the strong effects of dissipation. Our observation thus reveal the existence of a quasi-homogeneous regime at short-times and small-scales.
}


\section{Direct Estimation of Lagrangian Statistics}\label{sec:res}

In this section, we use the wind-tunnel Lagrangian dataset in order to extract two critical LSM parameters: the Lagrangian velocity decorrelation timescale and the Kolmogorov constant, $C_0$. To do so we calculate the autocorrelation and the structure functions and directly extract the two parameters from their definitions.

\subsection{Lagrangian Autocorrelation and Decorrelation Timescale}\label{sec:autocorrelation}

We estimated the Lagrangian autocorrelation functions using a formula equivalent to Eq.(2.6) in the paper by Guala et al.~\cite{Guala2007} with the ensemble averaging Eq.~\eqref{eq:mean}, as presented and discussed in Appendix~B.
Fig.~\ref{fig:RL_b3}(a) presents $\rho_{ii}$ for the three components of Lagrangian velocity in sub-volume b3. For all three components, a concave shape is seen at the origin, reminiscent of a parabolic decrease at small delay times, providing the way to estimate the Taylor micro-timescale~\cite{Tennekes1972}. The autocorrelation functions decrease monotonically with the increasing time lag, $\tau$. The rate of decrease is roughly the same for $\rho_{xx}$ and $\rho_{yy}$, whereas the decrease is faster for the vertical component, $\rho_{zz}$. These observations were robust throughout all of the sub-volumes and the two $Re_\infty$ cases.


The Lagrangian autocorrelation function $\rho_{ii}(\tau)$ does not decrease to zero within the range of our measurements, so we cannot use the integral in Eq.~\eqref{eq:autocorrelation} to estimate the $T_{L,i}$ directly. Therefore and similarly to previous Lagrangian measurements (Refs.~\cite{Mordant2004b, Ouellette2006}), we define a Lagrangian \textit{decorrelation timescale}, $T_i$, that we obtain by fitting our results to the autocorrelation function of an LSM. \revtwo{In accordance with the results of Section~\ref{sec:Quasi-homogeneity}, we use here a LSM for homogeneous flows.} \rev{Specifically, we used a least square minimization to fit our measurements to the autocorrelation function in Sawfords model, eq.~\eqref{eq:sawford_model}, to obtain $T_{1,i}$ and $T_{2,i}$ in each sub-volume and then define $T_i = T_{1,i} + T_{1,i}$.} It is noted that a calculation of $T_{L,i}$ using the full integral of $\rho_{ii}$ according to the theory, may result in a larger timescale on the order of the turnover timescale for large coherent structures above the canopy.
The concave shape of the autocorrelation function at $\tau\to 0$ implies that the exponential approximation according to first order models, e.g. Eq.~\eqref{eq:thomson_model}, is not compatible with our data. Instead, we used the second-order model by Sawford~\cite{Sawford1991}, Eq.~\eqref{eq:sawford_model}, that does take this concavity into account, similarly to Mordant et al.~\cite{Mordant2004b} and Ouellette et al.~\cite{Ouellette2006}.

We estimated $T_i$ for the three velocity components by fitting the data as shown in Fig.~\ref{fig:RL_b3}(a). The fit range was limited to the time lags that correspond to a half of the sub-volume crossing time, in order to avoid the possible finite volume effects (as discussed in Appendix~C and Ref.~\cite{Biferale2008}). This is a common approach in experimental data analysis because of increased uncertainty of correlations at larger time lags~\cite{George1978}.
Lastly, we note that every data point of $\rho_{ii}$ corresponds to the average of at least 15$\times10^3$ samples, where the relative mean squared error was of the order of a few percents.

\revtwo{It is worth noting that the Lagrangian velocity become decorrelated, namely $\rho_{ii}(\tau)$ reduced considerably, while the particles were still within the small sub-volumes we used. This is in agreement with an observation in our previous paper~\cite{Shnapp2019} where we detected the Taylor asymptotic dispersion regime ~\cite{Taylor1921}, and also with our observation in Section~\ref{sec:Quasi-homogeneity} of the quasi-homogeneous regime.}

\begin{figure*}[]
	\centering
	\begin{subfigure}[t]{0.5\textwidth}
		\centering
		\includegraphics[width=8cm]{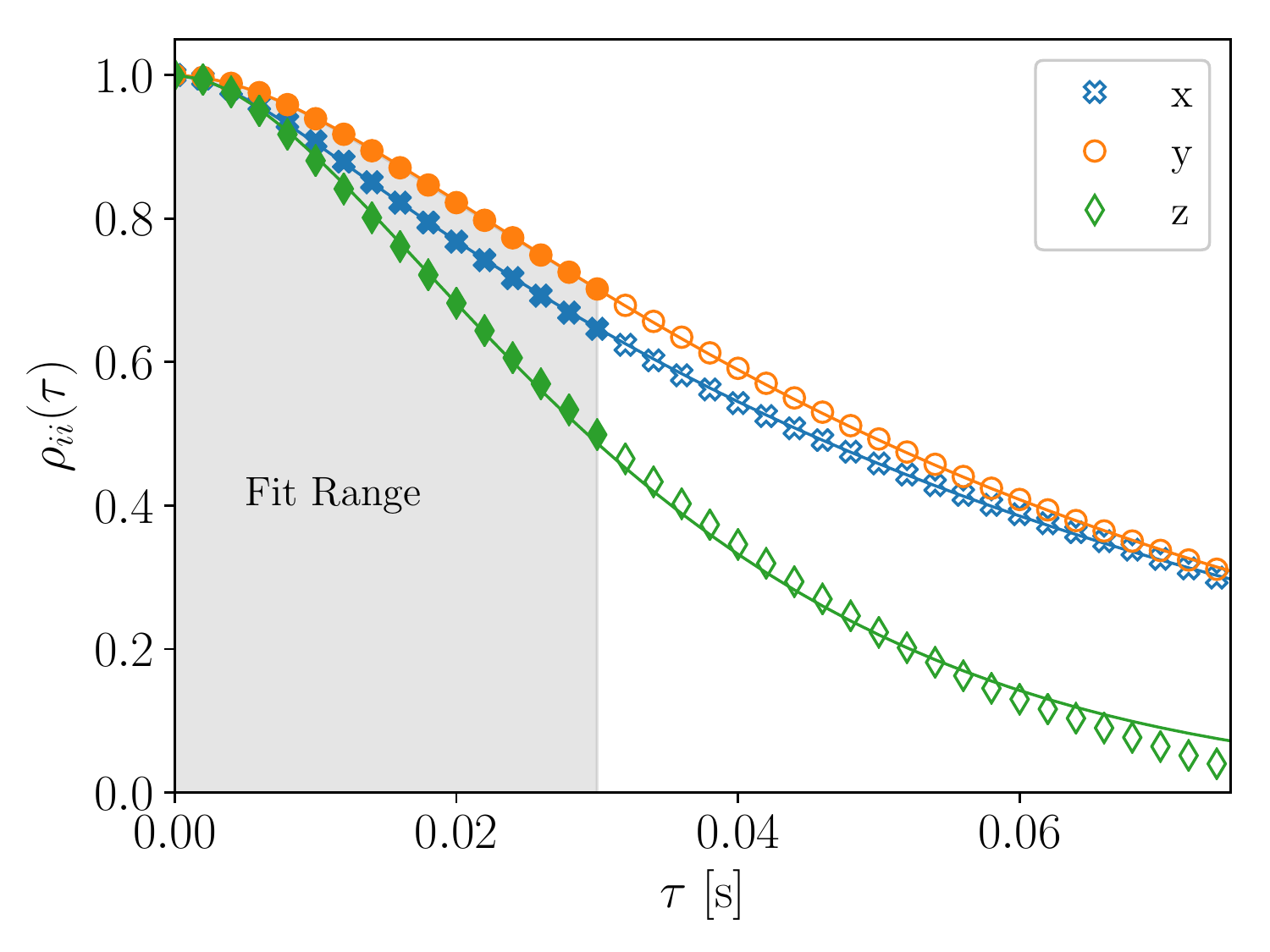}
		\caption{}
	\end{subfigure}%
	~ 
	\begin{subfigure}[t]{0.5\textwidth}
		\centering
		\includegraphics[width=8cm]{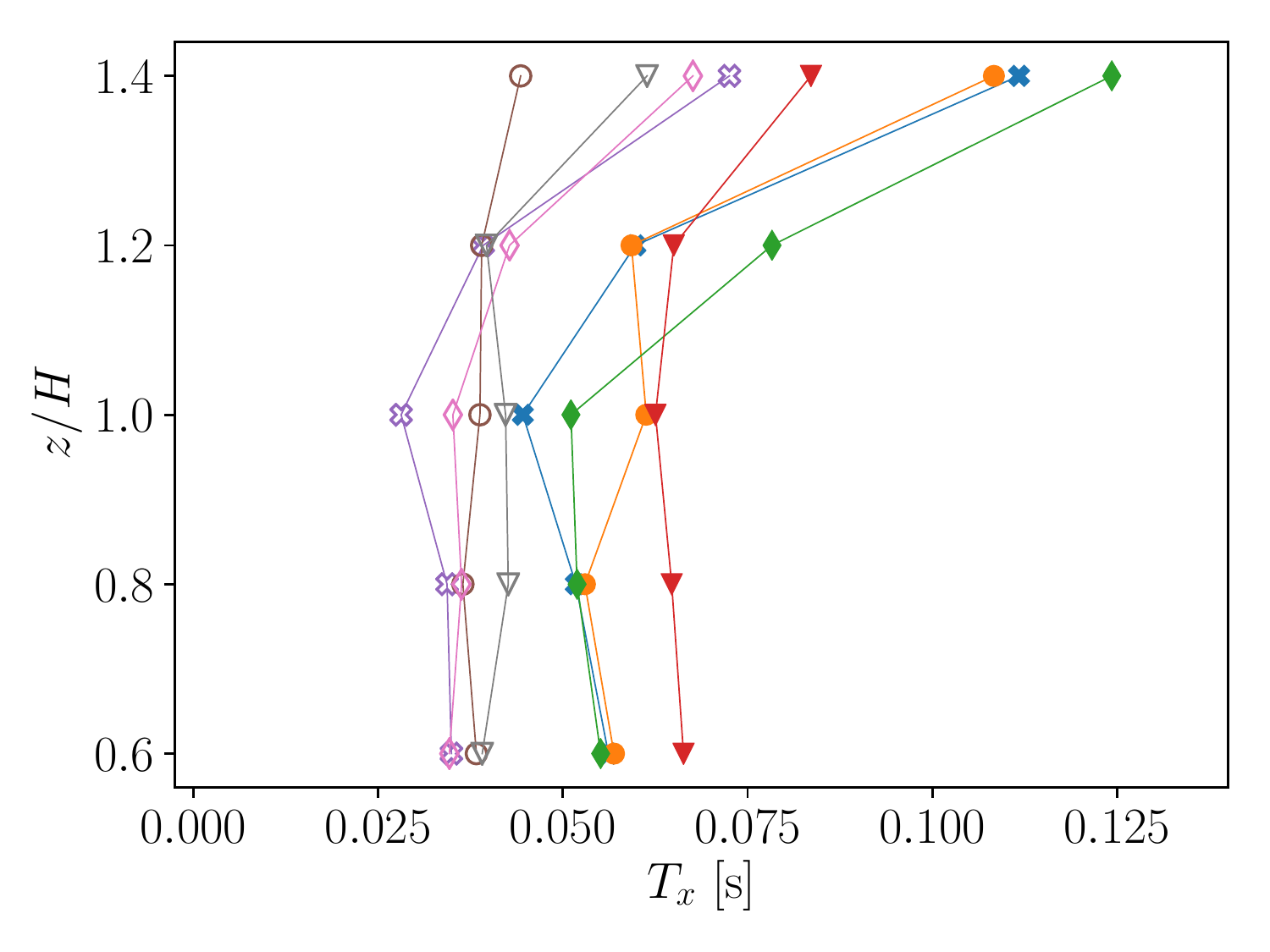}
		\caption{}
	\end{subfigure}%
	~\\
	\begin{subfigure}[t]{0.5\textwidth}
		\centering
		\includegraphics[width=8cm]{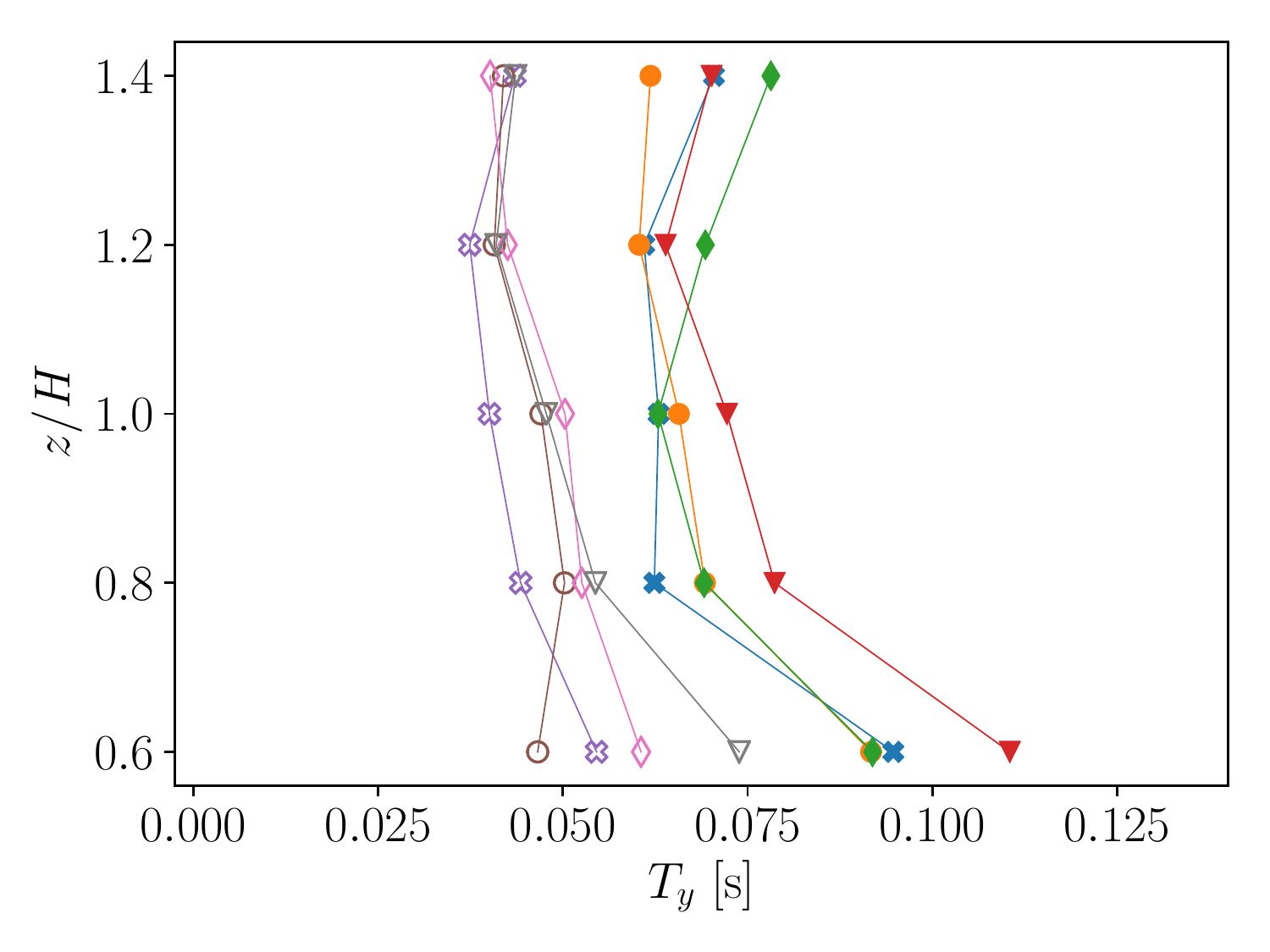}
		\caption{}
	\end{subfigure}%
	~ 
	\begin{subfigure}[t]{0.5\textwidth}
		\centering
		\includegraphics[width=8cm]{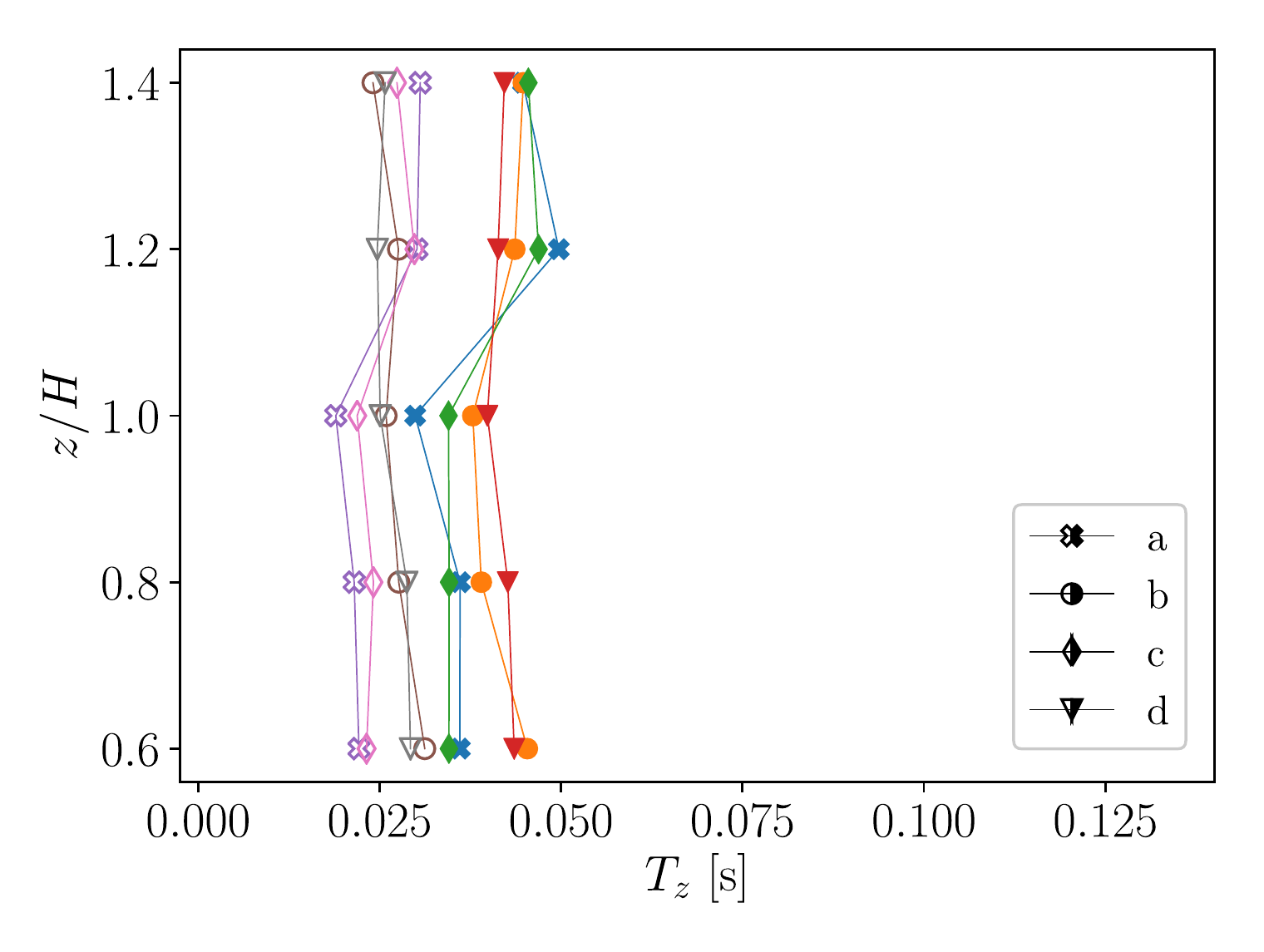}
		\caption{}
	\end{subfigure}
	\caption{(a) Autocorrelation function of Lagrangian velocity in the sub-volume $b3$ and $Re_\infty=16 \times 10^3$, presented against the time lag, shapes denote data points and lines represent fits to the model Eq.~\eqref{eq:sawford_model}. (b)-(d) The Lagrangian timescale for the $x$, $y$ and $z$ velocity component as a function of sub-volume height. Filled shapes stand for $Re_\infty=16\times10^3$ and empty shapes for $Re_\infty=26\times10^3$. \label{fig:RL_b3}}
\end{figure*}

The empirical Lagrangian decorrelation times, $T_i$, are presented in Fig.~\ref{fig:RL_b3}(b,c,d) for each velocity component as a function of height for the two Reynolds numbers tested. 
Inside the canopy layer, the values of $T_x$ are roughly constant and they increase above the canopy. In contrast, $T_y$ values are highest at the lowest sub-volume yet retain a roughly constant value above the canopy. The values of $T_z$ are the lowest of the three components and show only minor variation with height above $H$. Furthermore, the Lagrangian decorrelation times are consistently higher for the $Re_\infty=16\times10^3$ case, as compared to the higher $Re_\infty=26\times10^3$ case.

The distributions of $T_i$ can be associated with physical processes that are known to occur in canopy flows. The increase of $T_x$ above the canopy may be related to large scale coherent structures that are known to exist above canopies due to the shear instability, aka the mixing layer analogy~\cite{Raupach1996, Finnigan2000}. The increase of $T_y$ inside the canopy layer is attributed to the change in the roughness density with height -- the lower roughness elements caused an increased frontal area density and increased the "shielding" (e.g.~\cite{Vanderwel2019}), which contributed to a tunneling effect of a cross-flow inside the canopy layer. 
Lastly, lower values of $T_z$ as compared to $T_x$ and $T_y$ are in agreement with  estimations of the Eulerian integral timescales from velocity measurement, for instance by Refs.~\cite{Shaw1995, Poggi2008, DiBernardino2017}, which may be associated with an inclined orientation of coherent structures that was reported in the literature, e.g. by Shaw et al~\cite{Shaw1995} using two-point Eulerian correlations.


\subsection{Second Order Lagrangian Structure Function}\label{sec:Lag_Dii}

\begin{figure*}[]
	\centering
	\begin{subfigure}[t]{0.5\textwidth}
		\centering
		\begin{tikzpicture}
		\node[] (a) at (0,0) {\includegraphics[width=8cm]{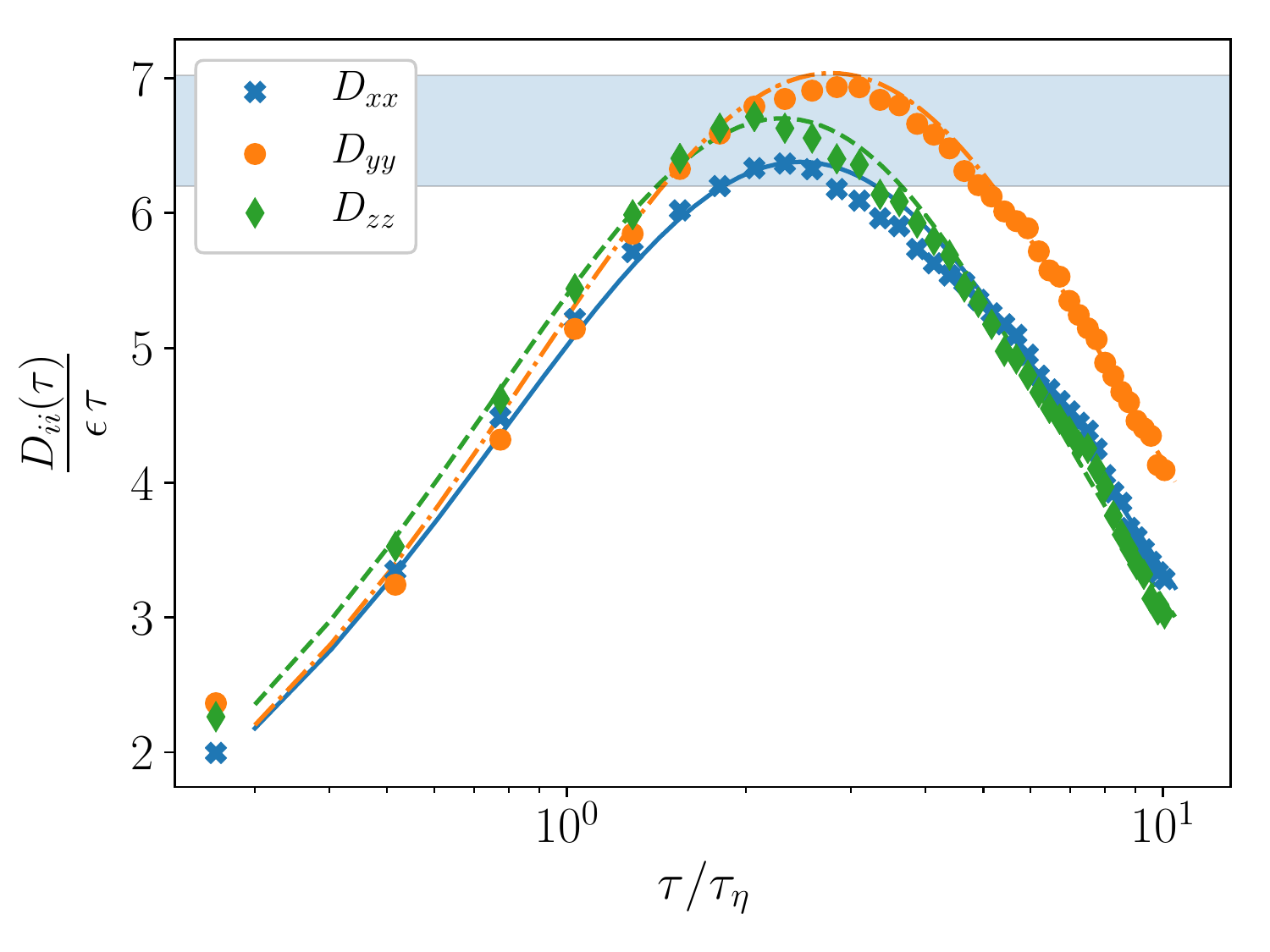}};
		\end{tikzpicture}
		\caption{}
	\end{subfigure}%
	~ 
	\begin{subfigure}[t]{0.5\textwidth}
		\centering
		\includegraphics[width=8cm]{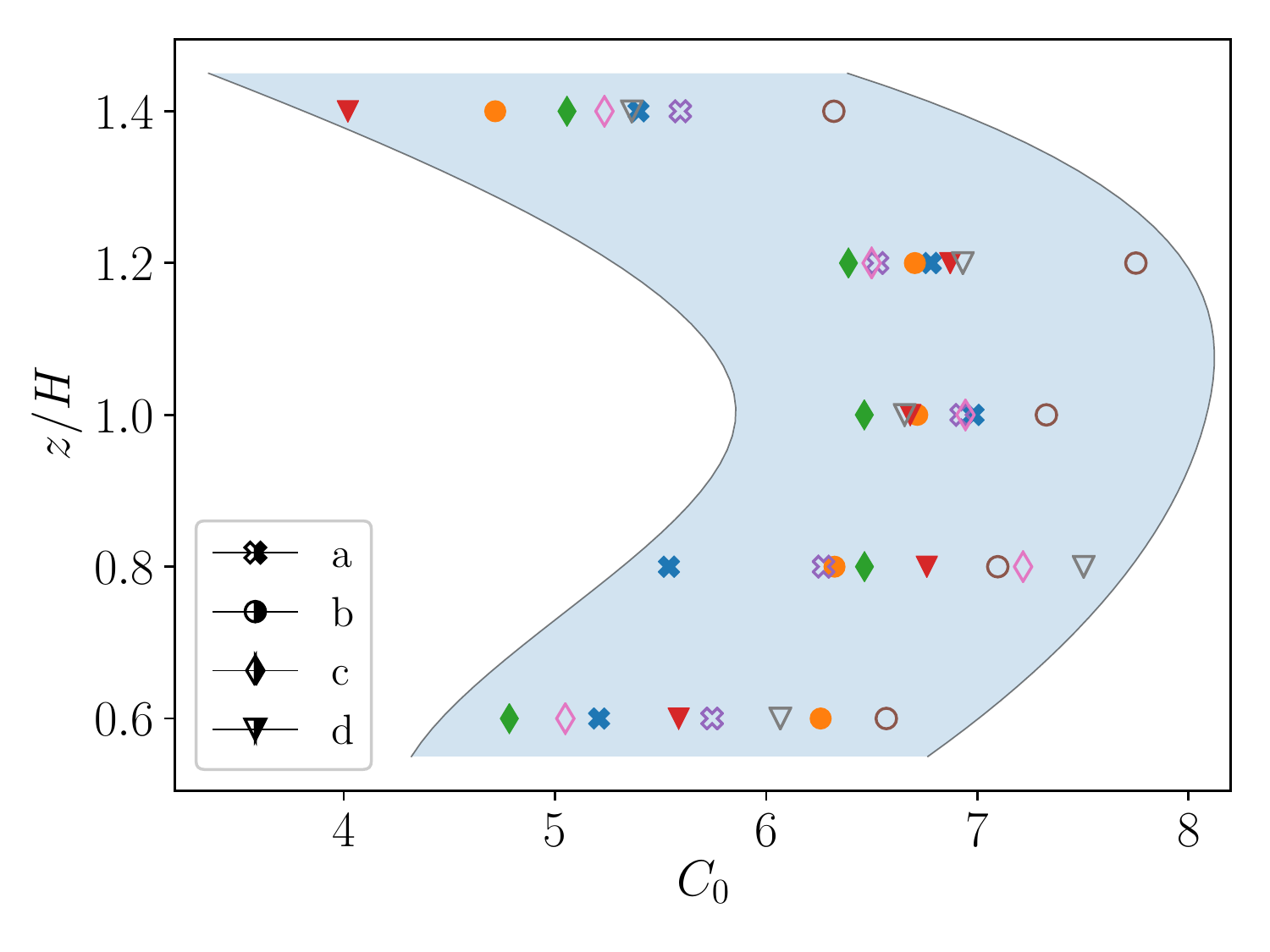}
		\caption{}
	\end{subfigure}
	\caption{(a) Second order Lagrangian structure function of velocity differences in sub-volume $b3$ and $Re=16 \times 10^3$, presented as a function of time normalized with the inertial-range scaling. \rev{The continuous, dashed, and dash-dotted lines represent Sawford's second order LSM~\cite{Sawford1991} fitted to the $x$, $y$ and $z$ components respectively}. (b) The estimated Kolmogorov constant Eq.~\eqref{eq:c0} from the $x$ velocity component, for all sub-volumes and two Reynolds numbers as a function of height. Full shapes for $Re_\infty=16\times 10^3$ and hollow for $Re_\infty=26\times 10^3$. \label{fig:Dll}} 
\end{figure*}

Using the sub-volume averaging, we estimated the Lagrangian second order structure function through its definition, Eq.~\eqref{eq:b}.
The results for trajectories in sub-volume b3 are shown in Fig.~\ref{fig:Dll}(a) in a compensated form, $D_{ii}/\epsilon \tau$, which corresponds to the Kolmogorov scaling in the inertial range.
For high Reynolds number HIT flows, the existence of Kolmogorov scaling in a Lagrangian inertial range would lead to a plateau in the compensated plot. The figure shows that such a plateau does not appear in our data, and instead, only narrow peaks are seen. Such peaks are characteristic of low to moderate Reynolds number flows, and similar observations were reported in numerous previous studies involving other types of flows, Refs.~\cite{Yeung1989, Mordant2001, Mordant2004b, Bec2006, Yeung2006a, Biferale2008, Toschi2009, Sawford2011, Yu2012}. Therefore, we use the typical empirical estimate of $C_{0,i}$, which is defined using the height of the peaks
\begin{equation}
C_{0,i} = \mathrm{max}_{\,\tau} \left[\frac{D_{ii}(\tau)}{ \epsilon \, \tau} \right].
\label{eq:c0}
\end{equation}
In the case of sub-volume b3, shown in Fig.~\ref{fig:Dll}(a), the values obtained are in the range $C_{0,i} \in (6.2, 7.0)$ for the three velocity components, meaning weak anisotropy $\sim \mathcal{O}(10\%)$.

We applied Eq.~\eqref{eq:c0} to the $D_{xx}(\tau)$ obtained in the different sub-volumes in the canopy flow model. The obtained values of $C_{0,x}$ are presented in Fig.~\ref{fig:Dll}(b) as a function of height. Values of $C_{0,i}$ calculated with $D_{yy}$ and $D_{zz}$ were very similar to Fig.~\ref{fig:Dll}(b) and are not shown for the sake of brevity. A considerable scatter is seen in the values, roughly $\sim 30\% $, that may be either due to horizontal inhomogeneity across sub-volumes through $Re_{\lambda}$ variations, or due to a sensitivity of Eq.~\eqref{eq:c0} to small uncertainties in the structure-function. In Fig.\ref{fig:Dll}(b), $C_{0,x}$ does not show a dependence on $Re_\infty$, which is consistent with the behavior of $Re_\lambda$. Going from inside the canopy and increasing in $z$, $C_{0,x}$ initially increases and roughly levels off at $C_{0,x}\approx 6.5$ in the range $1.0<z\leq1.2$. Further up above the canopy, $C_{0,x}$ decreased even though $Re_\lambda$ was seen to be highest at this height level.


\revtwo{The non-monotonous behavior of $C_{0,x}$ with $z$ suggests an important conclusion. In HIT, $C_0$ is governed by $Re_\lambda$ alone (e.g.  as suggested in \cite{Lien2002, Sawford2011}), however, the fact that here $C_0$ is not monotonous with $z$ while $Re_\lambda$ increases with $z$ monotonously suggests that in the canopy flow $C_{0}$ depends on other parameters in addition to $Re_\lambda$. An explanation for this observation can be offered through an analysis by Poggi et al.~\cite{Poggi2008}: the effect of wake production on the Lagrangian structure function, leading to scale dependence of the rate of dissipation in canopy flows (i.e. the spectral bump~\cite{Finnigan2000}), is "lumped" into $C_0$. Therefore, according to this explanation, the observed increase of $C_0$ at the top of the canopy in our measurements is due to a strong wake production, injecting turbulent kinetic energy at small scales.}


In Section~\ref{sec:autocorrelation} it was observed that the autocorrelation function was concave at the origin (Fig.~\ref{fig:RL_b3}), and that we obtained a good fit for the data using the second-order LSM, Eq.~\eqref{eq:sawford91}. In a homogeneous flow, the Lagrangian structure function and the $D_{ii}$ are simply related by
\begin{equation}
D_{ii}(\tau) = 2 \sigma_i^2 \left[ 1 - \rho_{ii}(\tau) \right]
\label{eq:D_ii_and_rho_ii} \,\, ,
\end{equation}
\revtwo{and thus with regard to the observed quasi-homogeneity of our flow it would be instructive to examine this relation here as well.} Thus, we used Eq.~\eqref{eq:D_ii_and_rho_ii} to fit the empirical data for $D_{ii}(\tau)$, where we used the expression for $\rho_{ii}(\tau)$ from the second order LSM, Eq.~\eqref{eq:sawford_model}. Specifically, we used a least square algorithm to fit the three model parameters for the three components of the structure function. The resulting curves are shown with lines in Fig.~\ref{fig:Dll}(a). The good match that was obtained for the empirical data shows that single-particle statistics in our measurements are represented well by Sawford's second order LSM~\cite{Sawford1991}, Eq.~\eqref{eq:sawford91}. \revtwo{This further reinforces the picture of quasi-homogeneity of Lagrangian statistics at short times in our flow.}



\section{Lagrangian Rapid Decorrelation}\label{sec:rapid_decorelation}

\subsection{Observation of Rapid Decorrelation in the Canopy Flow}

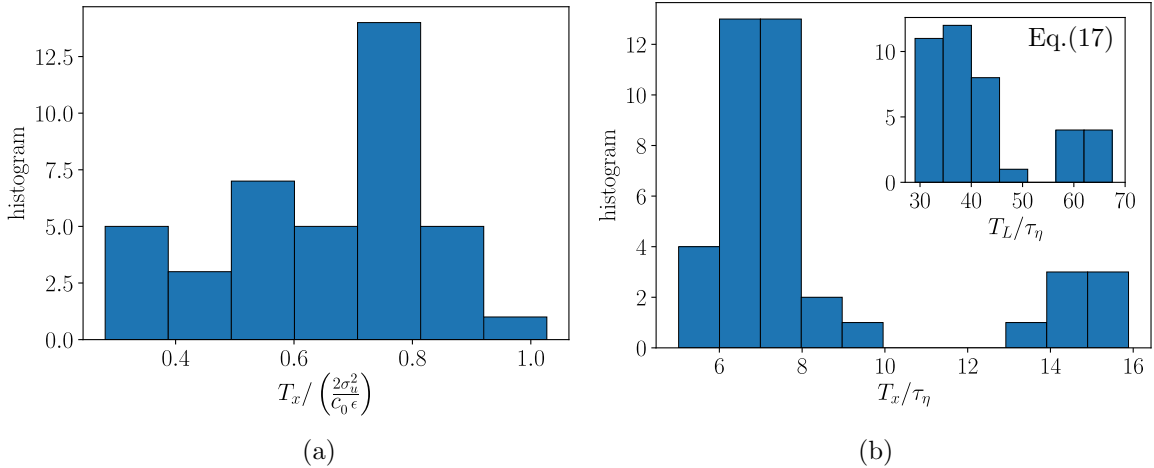
\begin{figure*}[]
	\centering
	\begin{tikzpicture}
	\node[] (a) at (-3.9,0) {\includegraphics[height=6.05cm]{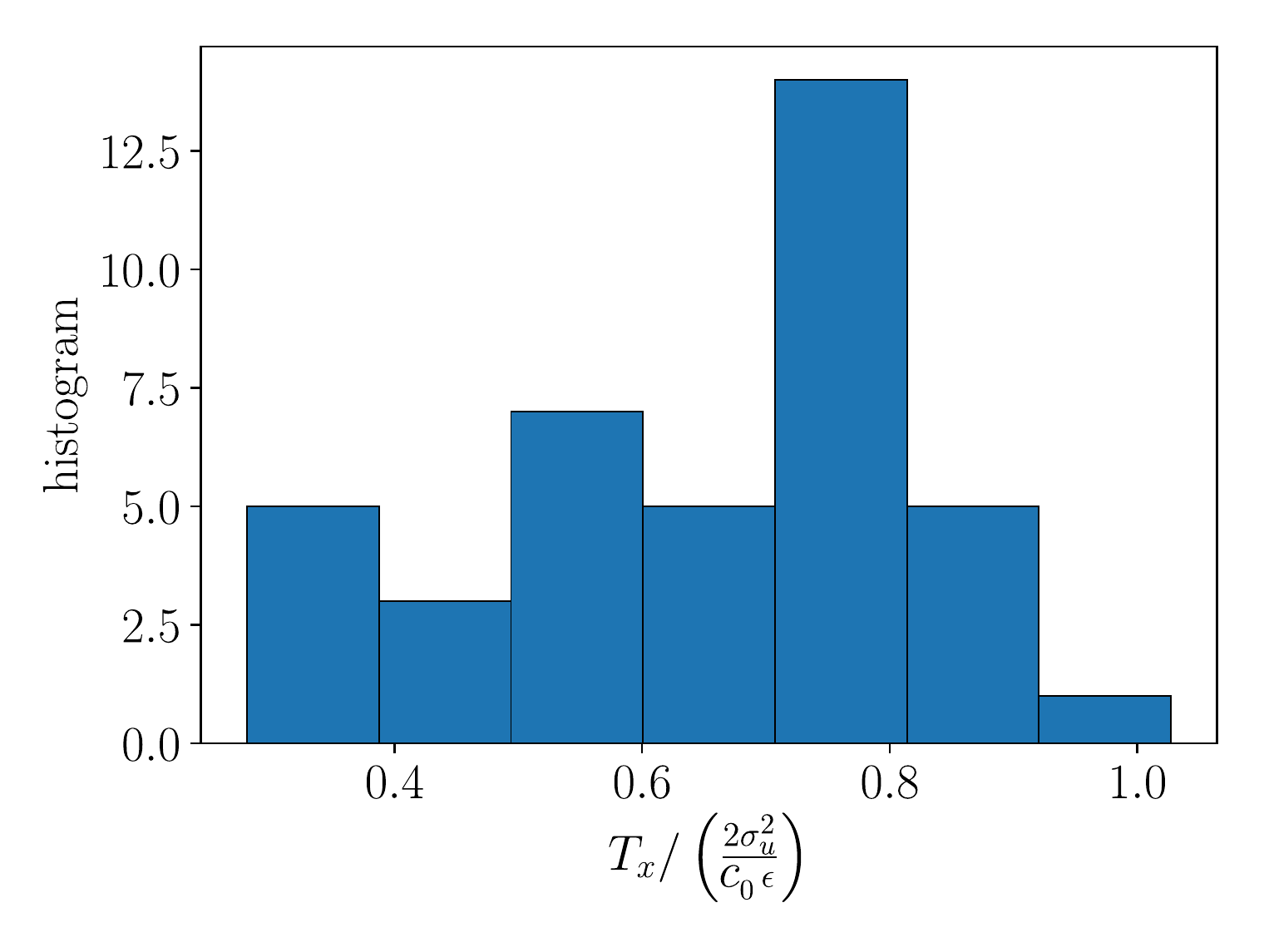}};
	\node[] (a) at (3.9,0.1) {\includegraphics[height=5.95cm]{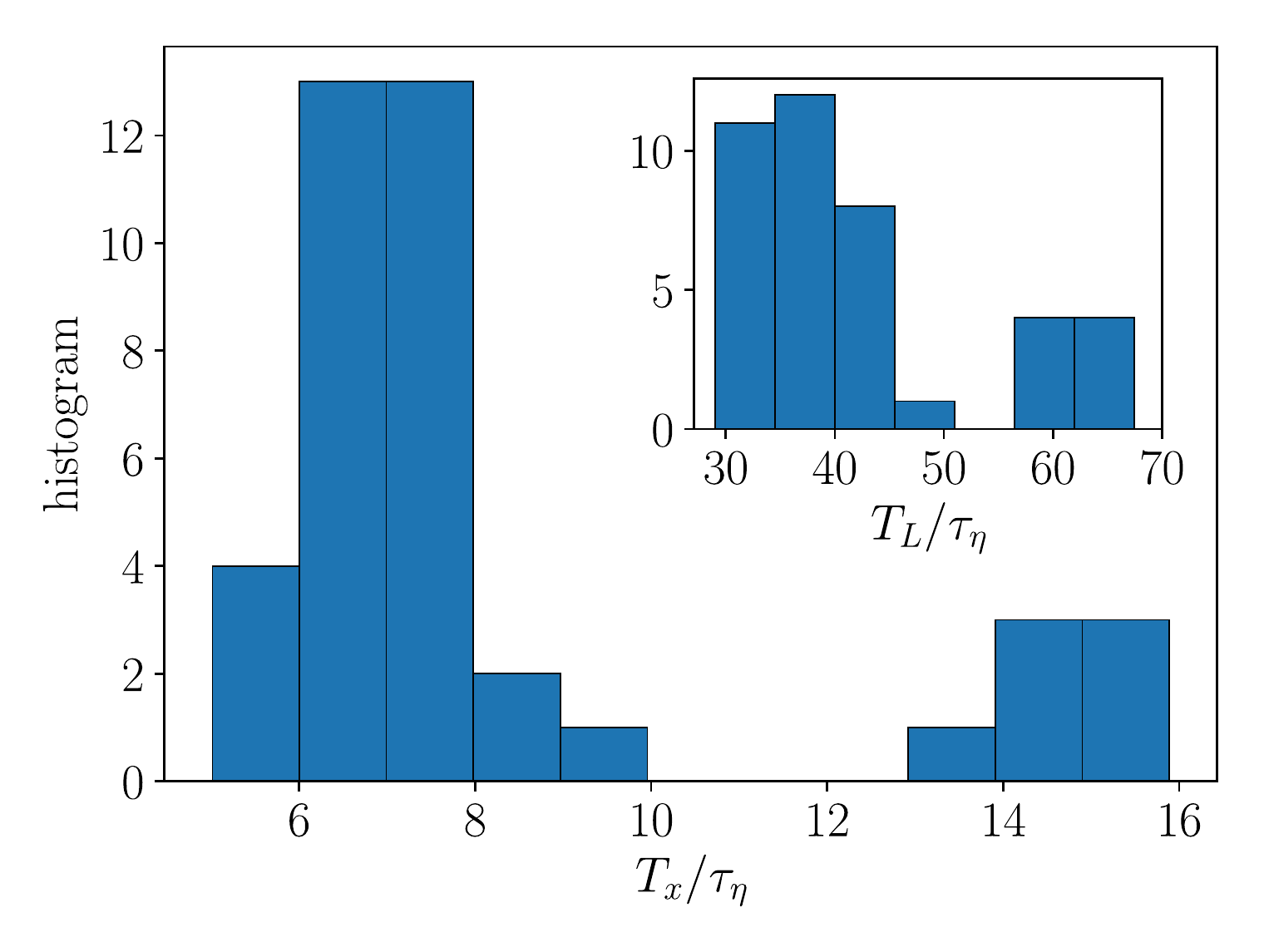}};
	\node[] (b) at (6.5,2.3) {Eq.\eqref{eq:Re_vs_Tt}};
	\node[] (b) at (3.9,-3.2) {(b)};
	\node[] (b) at (-3.5,-3.2) {(a)};
	\end{tikzpicture}
	\caption{Histograms of Lagrangian de-correlation times in the $x$ direction from all sub-volumes and two $Re_\infty$ (a total of 40 points), normalized with the (a) Eq.~\eqref{eq:T_x}, and (b) the dissipation timescale. \label{fig:TL_normed}}
\end{figure*}

In light of the small-scales' quasi-homogeneity, we find it instructive to compare our empirical estimates of $T_i$ with previous results for purely homogeneous turbulent flows. We first take the LSM Eq.~\eqref{eq:b}, according to which
\begin{equation}
T_{L,i} = \frac{2 v_i^{'2}}{C_{0,i} \epsilon} \,\,.
\label{eq:T_x}
\end{equation}
This relation was previously used to estimate the Lagrangian integral timescale based on Eulerian measurements in canopy flows, for example in Refs.~\cite{Aylor2001, Poggi2006a, Duman2016}. The comparison of our $T_i$ with Eq.~\eqref{eq:T_x} is shown in Fig.~\ref{fig:TL_normed}(a) through a histogram of the property $T_x / (\frac{2 \sigma_u^2}{C_0 \epsilon})$, taking values from all the sub-volumes and two $Re_\infty$.
The histogram shows a large scatter of values in the range [0.3--1] with an average of 0.65, and thus implies that the empirically estimated $T_{i}$ is significantly shorter than Eq.~\eqref{eq:T_x}.

The second comparison, in Fig.~\ref{fig:TL_normed}(b), compares the separation of scales $T_i / \tau_\eta$ in our canopy flow with that of HIT at similar $Re_\lambda$. 
A histogram of the decorrelation times, $T_x$, normalized with the dissipation timescale, $\tau_\eta$, is shown in Fig.~\ref{fig:TL_normed}(b). The results fall into two groups with values in the order of $T_x \sim 5-10 \tau_\eta$, and $T_x \sim 13-16 \tau_\eta$ that were seen to occur at different height levels, consistent with the increase of $Re_\lambda$ farther away from the wall. The values of $T_x/\tau_\eta$ seen in Fig.~\ref{fig:TL_normed}(b) are low as compared to values of $T_L / \tau_\eta$ usually encountered in HIT at comparable $Re_\lambda$. For example, Sawford et al.~\cite{Sawford2008} suggested
\begin{equation}
\frac{T_L}{\tau_\eta} = \left[ 4.77 + \left(\frac{Re_\lambda}{ 12.6} \right)^{4/3} \, \right]^{3/4}
\label{eq:Re_vs_Tt}
\end{equation}
following the empirical fit to DNS data. Using the $Re_\lambda$ values from our measurements, the estimates based on Eq.~\eqref{eq:Re_vs_Tt} are plotted in the inset of Fig.~\ref{fig:TL_normed}(b). Here the values are seen to be roughly an order of magnitude higher than those obtained by directly fitting the autocorrelation functions. Therefore, Fig.~\ref{fig:TL_normed} shows that the separation of scales in our canopy flow is much smaller than what would have been expected in comparable HIT case.

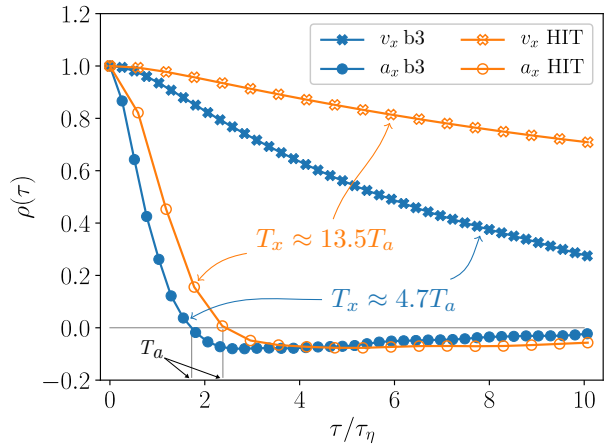
\begin{figure}[h]
	\centering
	\begin{tikzpicture}
	\node[] (a) at (0,0) {\includegraphics[width=8cm]{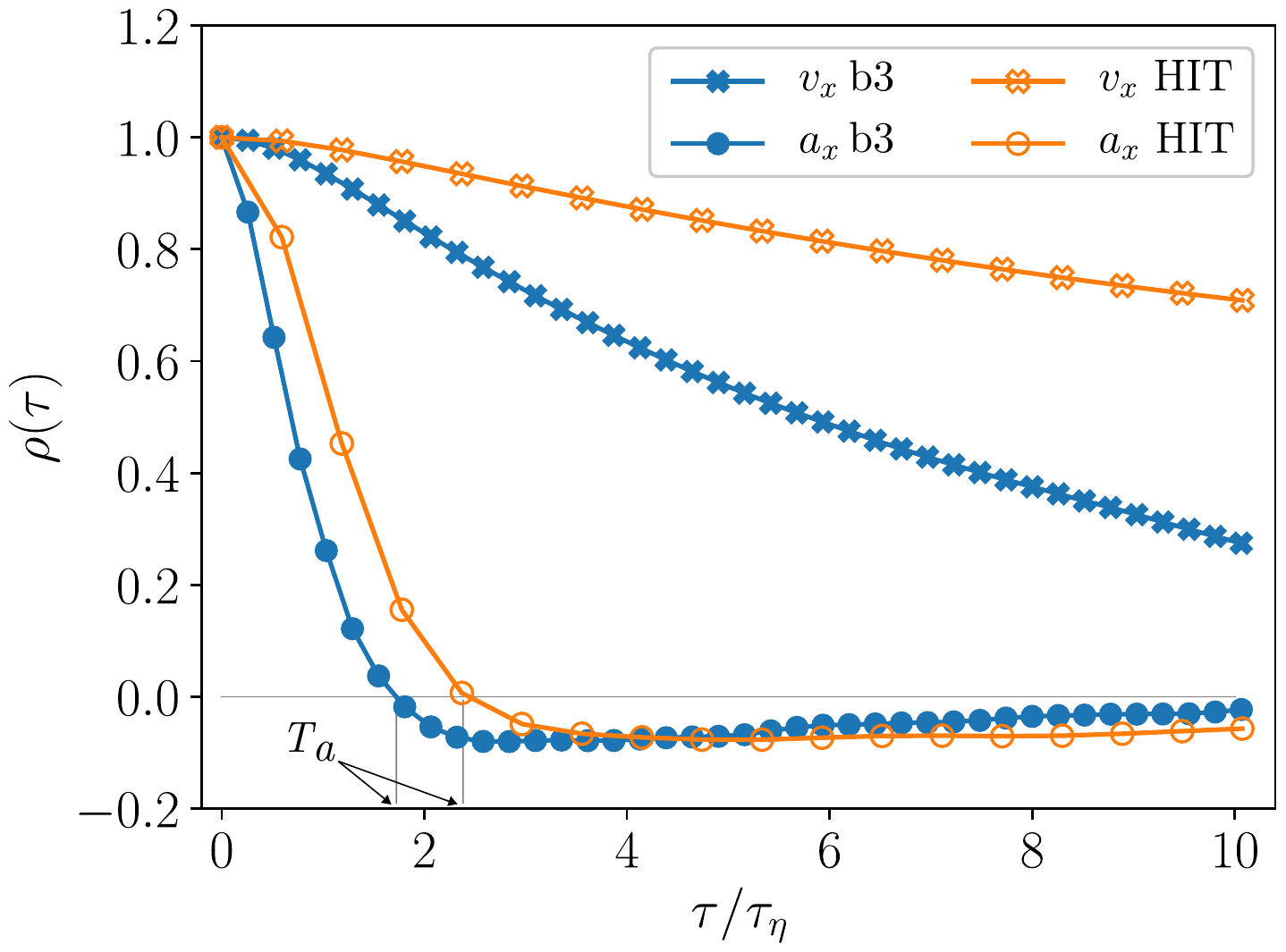}};
	\node[anchor=west,color=corange] (hit) at (-0.8,-.2) {$T_x\approx 13.5 T_a$};
	\draw[->,color=corange] (hit) to [out=70,in=-110] (1.1,1.35);
	\draw[->,color=corange] (hit) to [out=190,in=50] (-1.45,-0.75);
	
	\node[anchor=west,color=cblue] (b3) at (0.2,-1.) {$T_x\approx 4.7 T_a$};
	\draw[->,color=cblue] (b3) to [out=20,in=-110] (2.3,-0.15);
	\draw[->,color=cblue] (b3) to [out=180,in=30] (-1.55,-1.25);
	\end{tikzpicture}
	\caption{Lagrangian auto correlation functions of the acceleration component and the velocity component of Lagrangian particles. Hollow symbols correspond to trajectories for the DNS~\cite{Li2008, Yu2012,Shnapp_Liberzon:2018} and filled symbols to canopy trajectories from sub-volume b3 at $Re_\infty = 16\times10^3$. \label{fig:ax_vx_corr}}
\end{figure}

In Fig.~\ref{fig:TL_normed}, $T_i$ was compared with properties estimated using spatial information of the flow, namely $\epsilon$. We can reinforce the above observation of relatively small separation of scales by using a purely Lagrangian property that characterizes the small scales, such as the particles' acceleration, $\vec{a}=\frac{d \vec{v}}{dt}$. Thus, we contrast the autocorrelation of $v_i$ and $a_i$ in our canopy flow with those in a comparable HIT flow. For the HIT flow, we use the DNS data available from the Johns Hopkins Turbulence Database (JHTDB)~\cite{Li2008, Yu2012}. We downloaded this benchmark dataset in a previous study~\cite{Shnapp_Liberzon:2018}, and we use it here again to compare the autocorrelation functions. Autocorrelation functions from the DNS data and the data from sub-volume b3 are plotted together in Fig.~\ref{fig:ax_vx_corr}. Note that the $Re_\lambda$ in both cases are very similar, $Re_\lambda \approx 440$ in sub-volume b3 and $Re_\lambda \approx 433$ in the DNS~\cite{Yu2012}, and so they should present similar separation of scales if $T_L / \tau_\eta = f(Re_\lambda)$. Fig.~\ref{fig:ax_vx_corr} shows that the autocorrelation of the acceleration components decay at rates very similar to each other in both cases. However, the velocity in the canopy flow decorrelates much faster than in the HIT case. To be more quantitative, denoting by $T_a$ the time of the first zero crossing of the acceleration, it was found that $T_x / T_a \approx 4.7$ for the canopy data, while $T_x / T_a \approx 13.5$ for the HIT case. Therefore, Fig.~\ref{fig:ax_vx_corr} shows that while the $T_a$ was roughly the same in both cases, the velocity became decorrelated roughly 3 times faster in the canopy flow as compared to the HIT case. This observation reinforces the observation that the separation of timescales in our canopy flow is smaller than in the comparable HIT case.

Together, Figs.~\ref{fig:TL_normed} and~\ref{fig:ax_vx_corr} demonstrate that the Lagrangian velocity components in our canopy flow became decorrelated much faster than what would have been expected in a homogeneous isotropic turbulent flows at a similar Reynolds number. In particular, the above demonstrates that in canopy flows, unlike the HIT case, $T_i / \tau_\eta$ is \emph{not} a function of the Reynolds number alone. 
This observation will be termed in what follows \emph{rapid decorrelation}.

\subsection{Turbulence--Obstacle Interaction as the Source for Rapid Decorrelation}

In in section~\ref{sec:Quasi-homogeneity}, we observed a dominance of turbulent fluctuations over inhomogeneity in the Lagrangian dynamics, and in section~\ref{sec:Lag_Dii} we suggested that the structure function constant $C_0$ was affected by the wake production in accordance with the arguments of Poggi et al.~\cite{Poggi2008}. These observations show a strong influence of small scale dynamics on Lagrangian statistics in our measurements. These considerations lead us to put forth the notion that the observed rapid-decorrelation was also a consequence of wake effects due to the direct obstacle-flow interaction. In this section we examine this conjecture.

\begin{figure}[htbp]
	\centering
	\begin{subfigure}[t]{0.475\textwidth}
		\centering
		\includegraphics[width=8cm]{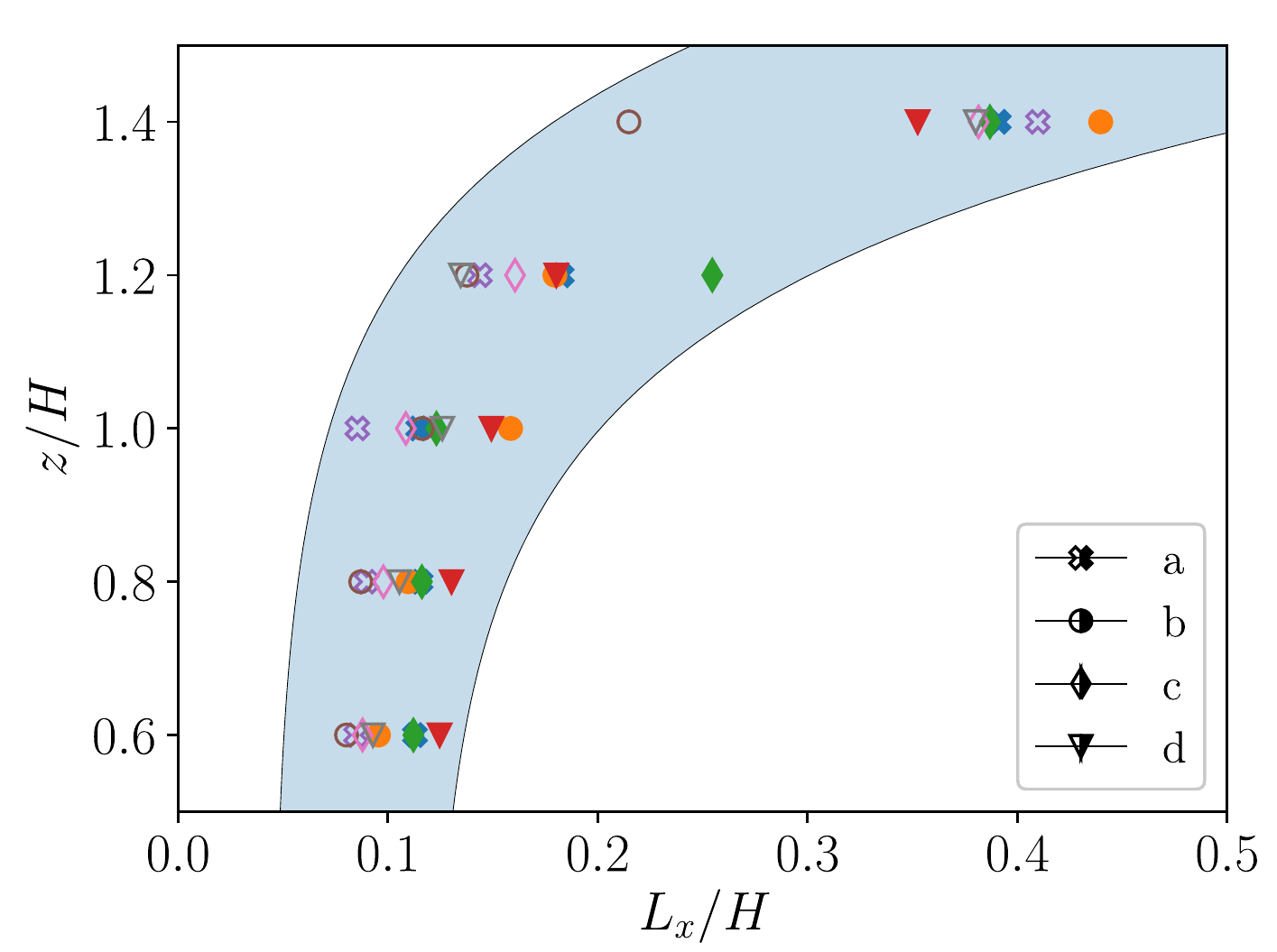}
		\caption{}
	\end{subfigure}
	\begin{subfigure}[t]{0.475\textwidth}
		\centering
		\includegraphics[width=7.3cm]{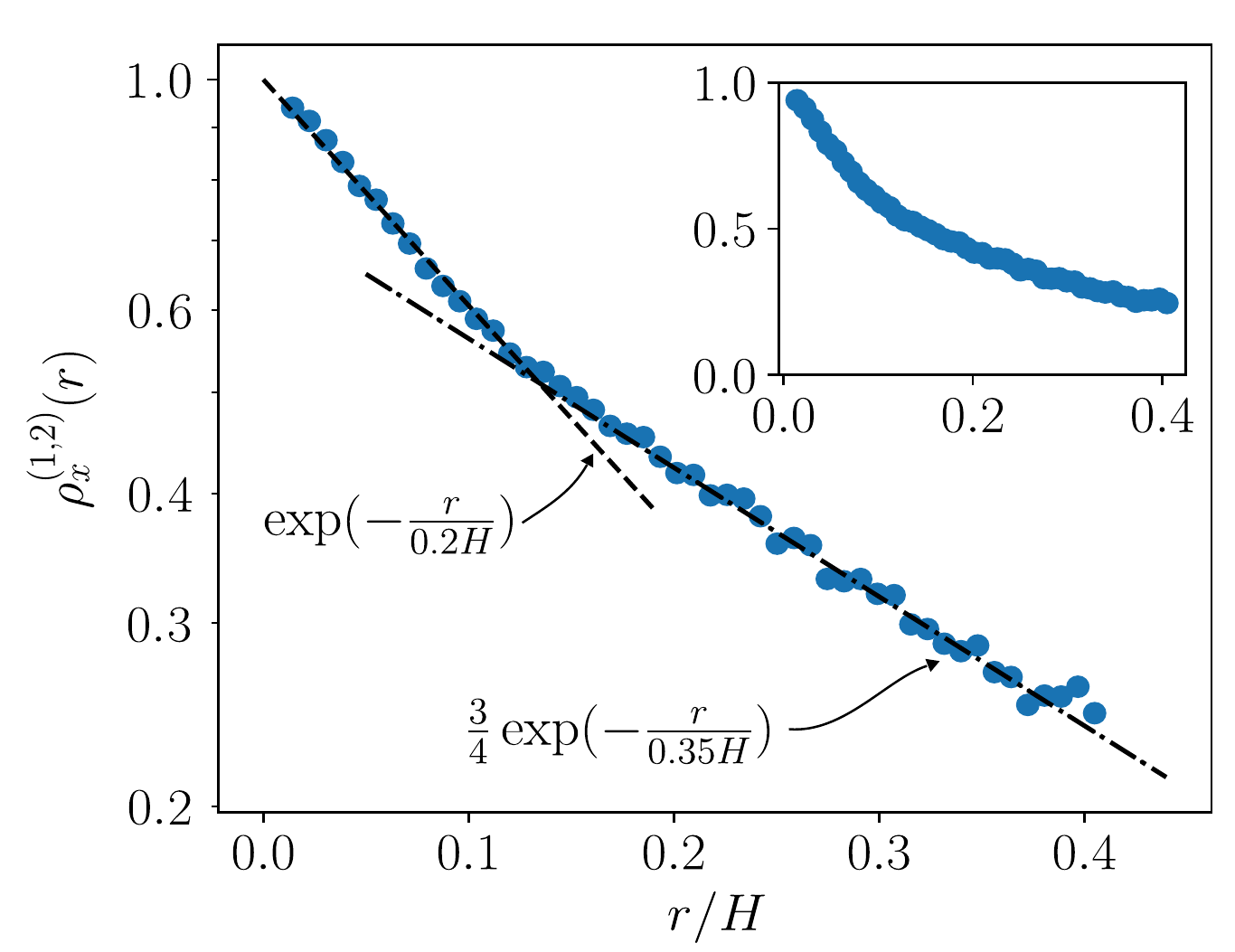}
		\caption{}
	\end{subfigure}
	\caption{(a) Lagrangian dispersion length scale $L_x = T_x\, \sigma_x$, presented as a function of height, both axes normalized by $H$. (b) Two-particle spatial correlation function, plotted against distance normalized by the canopy top height. Data for particles in sub-volume b3. \label{fig:Lagrange_length}}
\end{figure}

The encounter of the the flow with canopy obstacles leads to generation of drag that injects turbulent kinetic energy at flow scales with sizes that are determined by the geometry of the roughness obstacles, so-called wake-production~\cite{Finnigan2000, Poggi2004}. Therefore, if our conjecture was true, we would expect that the dispersion in the wakes will be dominated by flow disturbances with a similar size. The appropriate length scale of dispersion is $L_{i} = \sigma_i \, T_{L,i}$, which was observed to be correlated with the scale of forcing in previous experiments (Refs.~\cite{Mordant2004b, Xia2013}). The width of our obstacles was $d = 0.04 H$, so if the hypothesis is true we should see $L_i \sim d$. Thus, we calculated $L_x = T_x \, \sigma_x$ in the different sub-volumes and the results are presented in Fig.~\ref{fig:Lagrange_length}(a) against height. Inside the canopy layer, $z\leq H$,  $L_x$ is nearly constant, $L_x \approx 0.1 H = 2.5 d$ independently of $Re_\infty$, which may well be due to such wake disturbances. Above the layer, $L_x$ increases, reaching $\approx 0.35 H$ at the highest sub-volume. The lack of deviation of $L_x$ inside the canopy along with increase of $L_x$ at $z>H$ are consistent with the notion of weakening of the wake's influence above the canopy. Therefore, the estimated values of $L_x$ are consistent with the conjecture that rapid decorrelation occurred due to the flow disturbances in the obstacles' wakes.

To further support our conjecture, we wish to demonstrate that a disturbances occurred in our flow at the scale $L_x$. We demonstrate this through the two-particle spatial velocity correlation function. Specifically, let us define
\begin{equation}
\rho_{i}^{(1,2)}(r) = 
\frac{\av{ v_i'^{(1)}(t) \, v_i'^{(2)}(t) \,\, | \,\, r(t)}}
{ \text{Var}[v_i' (t)]_r}
\label{eq:spatial_correlation}
\end{equation}
where $v_i'^{(1)}$ and $v_i'^{(2)}$ are the velocity fluctuation of two different particles at the same time, and where the average in the numerator is performed for pairs of trajectories with a distance of $r(t) = |\vec{x}^{(1)}(t) - \vec{x}^{(2)}(t) |$ between them. In the denominator, $\text{Var}[v_i' (t)]_r$ is the variance of the velocity components calculated over the same ensemble of particles used in the numerator. Note that $\rho_{i}^{(1,2)}$ is a correlation with no separation in time but only in space, and thus it can be used to examine the spatial structure of the flow. Also note that $\rho_{i}^{(1,2)}(r)$ is analogous to the Eulerian two-point spatial velocity correlation function (for example see Shaw et al.~\cite{Shaw1995}), however, $\rho_{i}^{(1,2)}(r)$ is isotropic by construction since the average is performed over spherical shells. The two-particle correlation of $v'_x$, calculated using trajectories from sub-volume b3, is shown in Fig.~\ref{fig:Lagrange_length}(b). The same data is shown in linear-log scales in the main figure and in linear scales in the inset. The $\rho_{i}^{(1,2)}(r)$ decreases monotonously with $r$. In the range $r \lesssim 0.12 H$, the correlation decreases faster than at $r \gtrsim 0.12 H$. In the linear-log scales, the data at each interval roughly fits a straight line, where in each interval it has a different slope. Therefore, the data points were fitted with exponential decays with a different rate in each of these two ranges of $r$, which provided a good approximations of the data. The two fits, shown in dashed and dot-dashed lines, highlight the transition of the $\rho_{i}^{(1,2)}(r)$ from one rate of decay to another that occurs right at $r \approx 0.12 H \approx L_x$.
This transition of $\rho_{i}^{(1,2)}(r)$ from one rate of decay to another at $r\approx L_x$ may suggest an existence of flow disturbances of characteristic size $L_x\sim d$, namely corresponding to the width of the flow obstacles. Furthermore, such a transition of $\rho_{i}^{(1,2)}(r)$ at $r\approx L_x$ was robust for the sub-volumes inside the canopy, however not above it. Thus, since $T_x = L_x / \sigma_x$, we can speculate that such disturbances at the wake scale may have lead to the observed rapid-decorrelation; nevertheless we cannot prove this conjecture at this time. 

%


\revtwo{The two pieces of evidence presented above are in agreement with our conjecture, and thus leave the notion that wake production is the main cause of the rapid decorrelation of Lagrangian velocity a valid possibility. 
A conclusive proof will require further exploration, for example, by using a flow with various degrees of spectral short-circuiting versus inhomogeneity effects.}


\section{Discussion \& Conclusions}~\label{sec:conclusions}

\revtwo{In this work, we used experimental measurements to estimate Lagrangian statistics in a wind-tunnel canopy flow model directly in the Lagrangian framework. Our analysis indicates that turbulence-obstacle interaction, through wake-production had a significant effect on the short time Lagrangian statistics in our flow, and is relevant for Lagrangian stochastic models. In particular, our key result is that in spite of the large scale inhomogeneity (e.g. Figs~\ref{fig:RL_b3}, ~\ref{fig:Dll}(b), ~\ref{fig:TL_normed} and ~\ref{fig:Lagrange_length}(a)), we detected a quasi-homogeneous regime of Lagrangian statistics at short times and on the small scales. 
Furthermore, we show that the spatial variations of the separation of scales, $T_i/\tau_\eta$, and the Kolmogorov constant, $C_0$, cannot be explained by the variation of the Reynolds number, $Re_\lambda$; this suggests that unlike in HIT, they depend on additional parameters other than $Re_\lambda$.
The main difference is that the decorrelation timescale of the Lagrangian velocity is much shorter than would have been expected in a homogeneous case. We thus infer that both the rapid decorrelation and the alteration of $C_0$ are direct consequences of strong wake production.} 

\revtwo{The strong influence of the wakes on Lagrangian dynamics had important implications on our analysis. First, we found that the Lagrangian statistics in the quasi-homogeneous regime are recovered well by the second order LSM for homogeneous flows. Second, due to the small separation of scales, so-called rapid-decorrelation, we detected significant finite Reynolds number effects on the Lagrangian autocorrelation functions, and this is despite the fact that $Re_\lambda$ in our canopy flow was rather high, in the range of 350--850. Essentially, this is a demonstration that finite Reynolds number effects can be important in cases where the Richardson-Kolmogorov cascade is short-circuited.}

\revtwo{We expect that our results will be relevant for modeling short range dispersion in flows with intense dissipation and spectral short-circuiting (in particular, where $|\frac{1}{2}C_0 \epsilon R_{ij}^{-1} \, v'_j| \gg |\phi_i / g|$). 
We achieved this through high frontal area density (i.e. $\lambda_f = \sfrac{9}{16}$) leading to strong drag, thin obstacles that produced turbulent kinetic energy at a rather small scale, and obstacles with variable heights in consecutive rows.}

To conclude, the observations presented in this work make up a unique view on the Lagrangian dynamics in the canopy flows in the small scales. It is our view that short term dispersion modeling in canopy flows through LSMs may achieve increased accuracy by paying particular attention to wake dynamics in the canopy flows. Our work also highlights the importance of gathering Lagrangian statistics directly in the Lagrangian framework that is becoming possible with recently introduced technologies (i.e.~\cite{Shnapp2019} and references therein). Other important topics that were not dealt with here include the effects of mechanical diffusion, the mixing-layer analogy, and the inhomogeneity, on Lagrangian statistics, that due to the small scale of our observation could not have been assessed here, and thus leave considerable scope for future investigations.

\section*{Acknowledgement}

We are grateful to Meni Konn, Sabrina Kalenko, Gregory Gulitski, Valery Babin, Amos Shick and Mordechai Hotovely, for their assistance in preparing and performing the wind tunnel experiment. \revtwo{We also thank an anonymous reviewer for provoking an important discussion.} This study is supported by the PAZY grant number
2403170.

\bibliography{bibliography}

\section*{Appendix A - sub-volume flow parameters}\label{app1}

\begin{figure}[H]
	\centering
	\includegraphics[width=7cm]{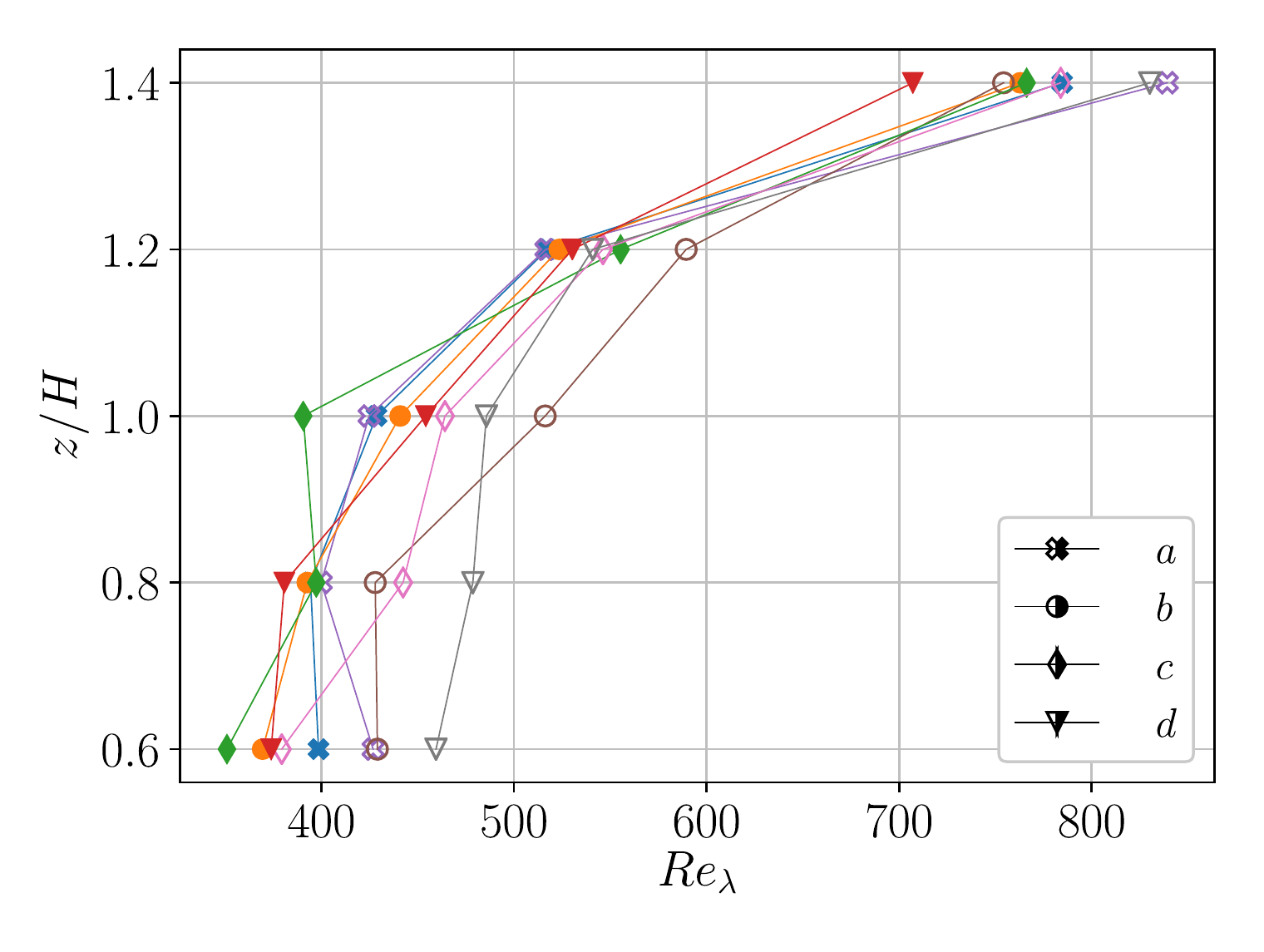}
	\caption{Change of the Taylor microscale Reynolds number with height for the various sub-volumes. full symbols are for $Re_\infty=1.6\times10^3$, and open symbols are for $Re_\infty=2.6\times10^3$. \label{fig:Re_lambda}}
\end{figure}

The two following tables present the values of the sub-volume parameters that were estimated according to the description in Section~\ref{sec:dissipation}. The change of $Re_\lambda$ with height is also presented in Fig.~\ref{fig:Re_lambda}.

\begin{table}[ht!]
	\scriptsize
	\centering
	\begin{ruledtabular}
		\begin{tabular}{ l c c c c c c c }
			$sv$ & $\tilde{u}$ [m/s] & $\epsilon$ [W/kg] & $\eta$ [mm] & $\tau_\eta$ [s] & $\lambda$ [mm] & $Re_\lambda$ & $H/\eta$ \\ \hline
			$a1$ & 0.42 & 0.201 & 0.36 & 0.009 & 14.14 &  398 & 277 \\
			$a2$ & 0.45 & 0.256 & 0.34 & 0.008 & 13.25 &  394 & 295 \\
			$a3$ & 0.49 & 0.304 & 0.32 & 0.007 & 13.23 &  428 & 308 \\
			$a4$ & 0.51 & 0.244 & 0.34 & 0.008 & 15.35 &  517 & 291 \\
			$a5$ & 0.62 & 0.239 & 0.34 & 0.008 & 19.01 &  784 & 290 \\
			$b1$ & 0.36 & 0.123 & 0.41 & 0.011 & 15.41 &  369 & 245 \\
			$b2$ & 0.42 & 0.193 & 0.36 & 0.009 & 14.18 &  392 & 275 \\
			$b3$ & 0.47 & 0.250 & 0.34 & 0.008 & 14.09 &  440 & 293 \\
			$b4$ & 0.50 & 0.233 & 0.35 & 0.008 & 15.63 &  523 & 288 \\
			$b5$ & 0.65 & 0.305 & 0.32 & 0.007 & 17.63 &  762 & 308 \\
			$c1$ & 0.42 & 0.248 & 0.34 & 0.008 & 12.59 &  350 & 292 \\
			$c2$ & 0.43 & 0.210 & 0.36 & 0.008 & 13.97 &  397 & 280 \\
			$c3$ & 0.46 & 0.286 & 0.33 & 0.007 & 12.82 &  390 & 303 \\
			$c4$ & 0.52 & 0.231 & 0.35 & 0.008 & 16.13 &  555 & 287 \\
			$c5$ & 0.62 & 0.245 & 0.34 & 0.008 & 18.66 &  766 & 291 \\
			$d1$ & 0.40 & 0.175 & 0.37 & 0.009 & 14.18 &  373 & 268 \\
			$d2$ & 0.40 & 0.175 & 0.37 & 0.009 & 14.32 &  380 & 268 \\
			$d3$ & 0.47 & 0.229 & 0.35 & 0.008 & 14.61 &  454 & 287 \\
			$d4$ & 0.50 & 0.218 & 0.35 & 0.008 & 15.99 &  530 & 283 \\
			$d5$ & 0.66 & 0.377 & 0.31 & 0.006 & 16.10 &  707 & 325 \\
		\end{tabular}
	\end{ruledtabular}
	\caption{Turbulence parameters for each sub-volume for the $Re_\infty = 16\times 10^3$ case.} 
\end{table}

\begin{table}[ht!]
	\scriptsize
	\centering
	\begin{ruledtabular}
		\begin{tabular}{ l  c  c  c  c  c  c  c }
			$sv$ & $\tilde{u}$ [m/s] & $\epsilon$ [W/kg] & $\eta$ [mm] & $\tau_\eta$ [s] & $\lambda$ [mm] & $Re_\lambda$ & $H/\eta$ \\
			\hline
			$a1$ & 0.53 & 0.422 & 0.30 & 0.006 & 12.16 &  426 & 334 \\
			$a2$ & 0.54 & 0.551 & 0.28 & 0.005 & 11.01 &  399 & 357 \\
			$a3$ & 0.60 & 0.703 & 0.26 & 0.005 & 10.68 &  424 & 379 \\
			$a4$ & 0.64 & 0.611 & 0.27 & 0.005 & 12.19 &  516 & 366 \\
			$a5$ & 0.83 & 0.669 & 0.27 & 0.005 & 15.19 &  839 & 375 \\
			$b1$ & 0.47 & 0.257 & 0.34 & 0.008 & 13.80 &  429 & 295 \\
			$b2$ & 0.50 & 0.352 & 0.31 & 0.007 & 12.74 &  427 & 319 \\
			$b3$ & 0.60 & 0.497 & 0.29 & 0.005 & 12.83 &  516 & 348 \\
			$b4$ & 0.64 & 0.487 & 0.29 & 0.006 & 13.79 &  589 & 346 \\
			$b5$ & 0.75 & 0.544 & 0.28 & 0.005 & 15.17 &  754 & 356 \\
			$c1$ & 0.52 & 0.490 & 0.29 & 0.006 & 11.05 &  379 & 347 \\
			$c2$ & 0.53 & 0.412 & 0.30 & 0.006 & 12.46 &  442 & 332 \\
			$c3$ & 0.60 & 0.587 & 0.28 & 0.005 & 11.68 &  464 & 363 \\
			$c4$ & 0.64 & 0.565 & 0.28 & 0.005 & 12.79 &  546 & 359 \\
			$c5$ & 0.81 & 0.706 & 0.26 & 0.005 & 14.49 &  783 & 380 \\
			$d1$ & 0.53 & 0.371 & 0.31 & 0.006 & 13.03 &  459 & 323 \\
			$d2$ & 0.52 & 0.327 & 0.32 & 0.007 & 13.72 &  478 & 313 \\
			$d3$ & 0.59 & 0.524 & 0.28 & 0.005 & 12.28 &  485 & 353 \\
			$d4$ & 0.63 & 0.528 & 0.28 & 0.005 & 12.94 &  540 & 353 \\
			$d5$ & 0.88 & 0.876 & 0.25 & 0.004 & 14.13 &  830 & 401 \\
		\end{tabular}
	\end{ruledtabular}
	\caption{Turbulence parameters for each sub-volume for the $Re_\infty = 26\times 10^3$ case.} 
\end{table}

\section*{Appendix B - Empirical Estimation of Autocorrelation Functions}\label{App:C}
The autocorrelation of the random signals in this work were calculated as follows. Consider the set of $i=1\,\dots\,N$ random series samples $a_i(\tau)$ of the random variable $a(\tau)$. Generally speaking, the average of $a(\tau)$ and its standard deviation may change with $\tau$, where$\tau = t-t_0$, and $t_0$ is the time at which the record of $a_i$ began). The average of $a(\tau)$ is defined as
\begin{equation}
\mu(\tau) = \frac{1}{N} \sum_{i=1}^N a_i(\tau)
\end{equation}
the fluctuations relative to the average are denoted $a_i(\tau)' = a_i(\tau) - \mu(\tau)$, and the standard deviation of $a(\tau)$ is defined
\begin{equation}
\sigma(\tau) = \left[ \frac{1}{N} \sum_{i=1}^N a_i'^2(\tau) \, \right]^{1/2} \,\, .
\label{eq:avg_std}
\end{equation}
Note that these two definitions correspond to the sub-volume average introduced in section~\ref{sec:sub_volume_approach}.
Then the autocorrelation of $a$ is calculated as follows
\begin{equation}
\rho(\tau) = 
\frac{ \frac{1}{N} \sum_{i=1}^N [ a_i'(0) \, a_i'(\tau) ]}{\sigma(0) \, \sigma(\tau)}
\label{eq:autocorrelation_estimation}
\end{equation}

The estimator Eq. (D.3) uses the average and the standard deviations that are allowed to change with $\tau$. As discussed by Guala et al.~\cite{Guala2007}, the Lagrangian trajectories with long tracking time possibly belong to a subset of ``weak turbulence''. Consequently, an estimator of $\rho$ that uses a single value $\mu$ and $\sigma$ averaged over all values of $\tau$ is a biased estimator that may under predicts $\rho(\tau)$ at long times. Therefore using the definition $\sigma(\tau)$, that changes with $\tau$ in Eq.(D.3), prevents this underestimation at long time lags. This issue was discussed in details by Guala et al.~\cite{Guala2007}, where the biased estimator in their paper was denoted Eq. (2.1), and the unbiased estimator Eq. (D.3) here is equivalent to their Eq.(2.6).

\bigskip

\begin{figure}[H]
	\centering
	\includegraphics[width=7cm]{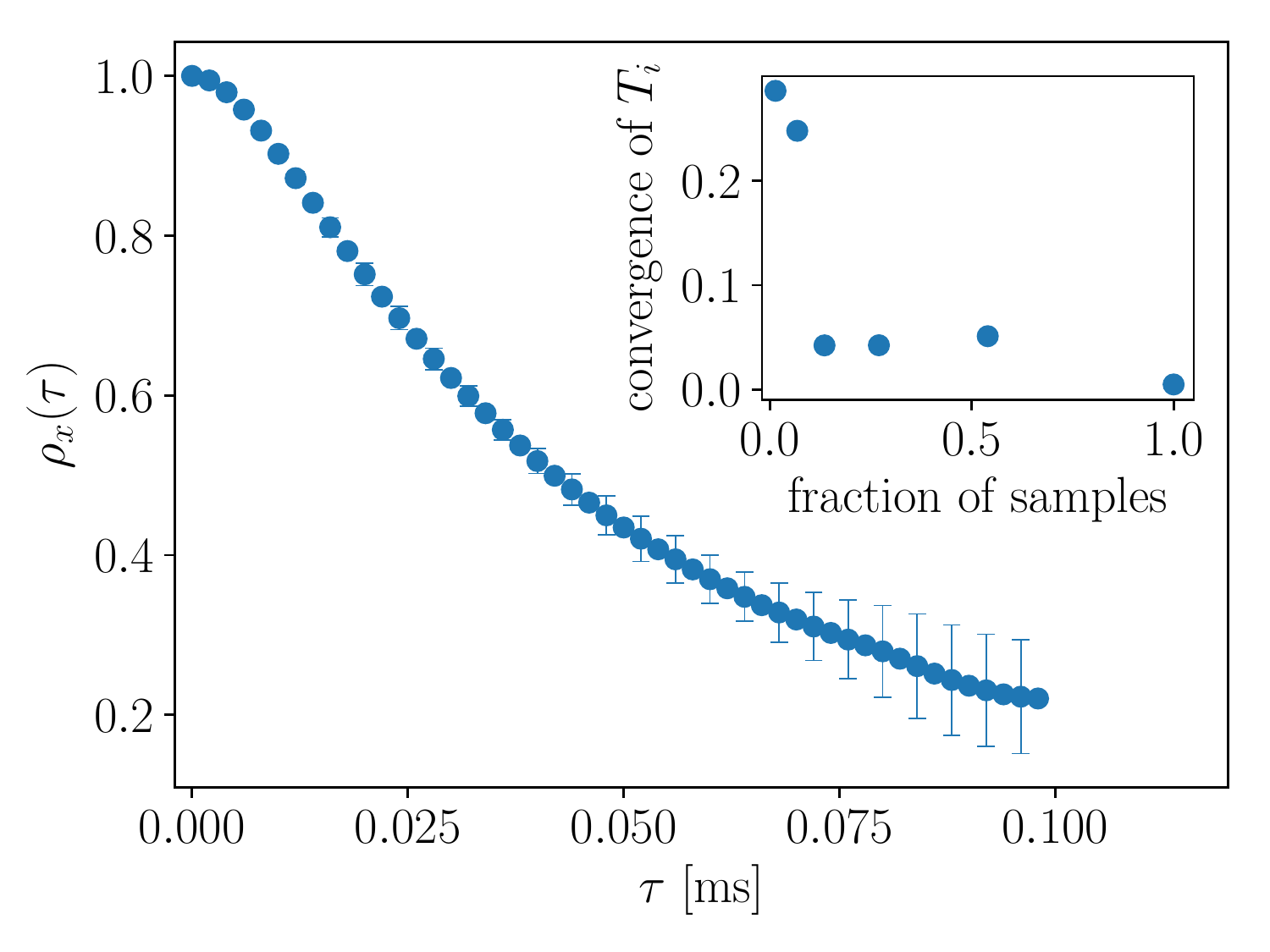}
	\caption{Convergence plots for the autocorrelation function (main panel) and for the decorrelation timescale (inset) for trajectories in sub-volume b3. The error bars in the main panel represent the range of scatter for $\rho_{xx}$ calculated using 3 subsamples of the data. The inset shows the relative error in estimating $T_i$ using subsamples of different sizes. \label{fig:rho_convergence}}
\end{figure}

The autocorrelation functions were calculated in this article using many samples that were measured during the long experimental runs we have conducted, $\sim 12--15$ minutes each. To demonstrate that our estimations of the autocorrelation function and the decorrelation timescale are converged we show in the main panel of Fig.~\ref{fig:rho_convergence} the Lagrangian autocorrelation function with error bars that represent the results of a bootstrapping calculation. Specifically, the dataset of trajectories in sub-volume b3 were divided to three groups and the autocorrelation $\rho_{xx}$ was calculated separately. The error bars show the range of scatter of the results for the three groups and represents a small degree of uncertainty in the range relevant for our study. Furthermore, the inset shows the convergence of $T_i$ that was calculated using subsamples of our data with different sizes. The relative error of $T_i$ is seen to decrease rapidly with the subsample size. Therefore, Fig.~\ref{fig:rho_convergence} demonstrates that the autocorrelation and the decorrelation times were converged in our experiment, and suggests an uncertainty of up to a few percents.

\section*{Appendix C - finite volume effect on correlation}

\begin{figure}[H]
	\centering
	\includegraphics[width=7cm]{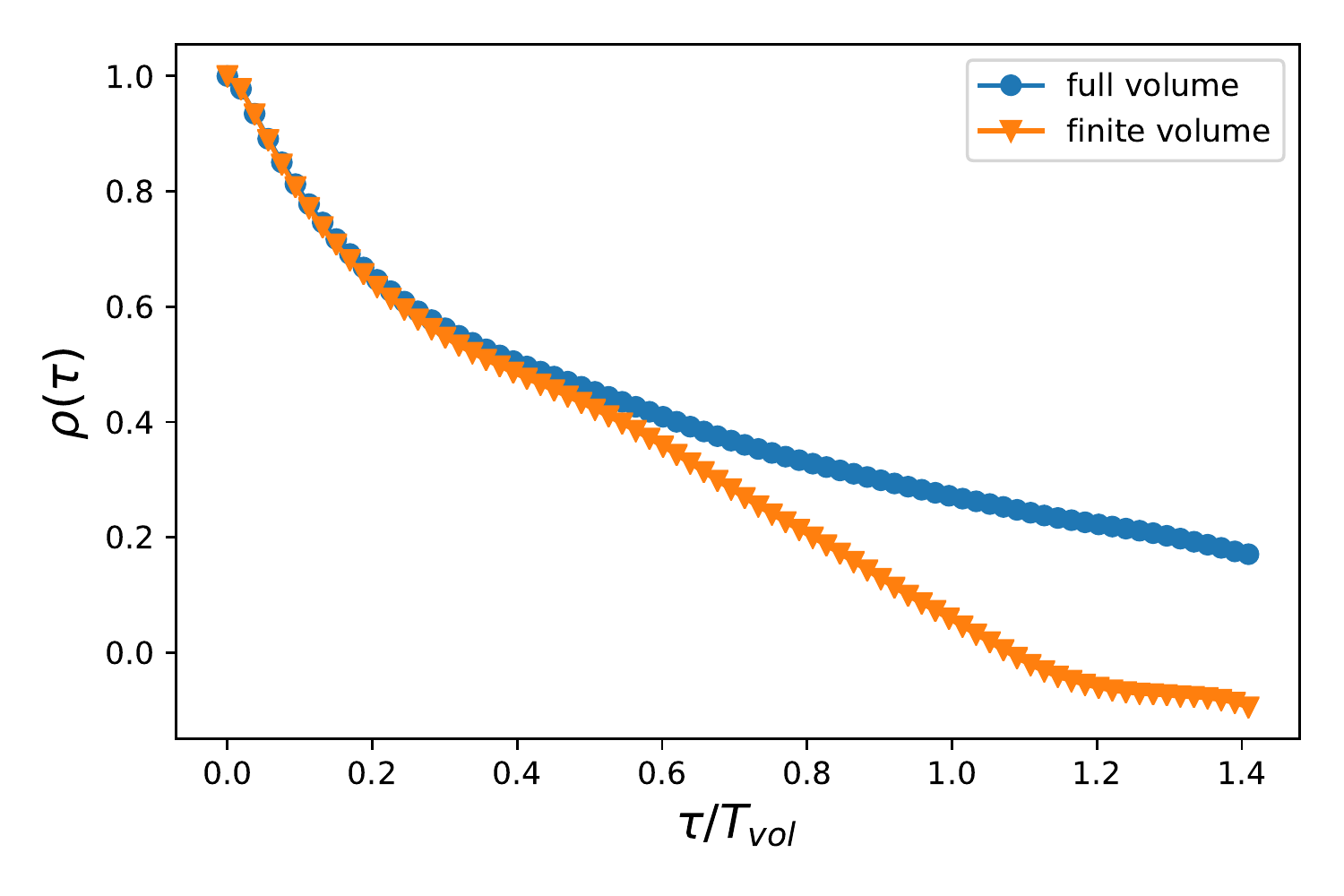}
	\caption{Lagrangian velocity autocorrelation over the full data set compared with trajectories that were truncated to be within finite volume. Plotted against time normalized by the volume timescale, Eq.~\eqref{eq:Tvol}. \label{fig:R_finite_vol}}
\end{figure}

Since the volume of observation is finite and due to the fact that occupation times of particles within the finite volumes are dependent on their velocity, a natural bias occurs in the estimation of Lagrangian velocity autocorrelation functions in PTV experiments. To minimize this effect on the results shown in this work, the estimation of Lagrangian timescales was performed in this work only on short times, such that most of the particles do not have sufficient time to leave the observation volume. A time scale for the occupation times within a volume of dimension $L$ is    
\begin{equation}
T_{vol} = \frac{L}{u'}
\label{eq:Tvol}
\end{equation}
with $u'$ being the root mean squared value of particle velocities. In Fig.~\ref{fig:R_finite_vol}, the Lagrangian autocorrelation function for the $x$ velocity component is presented against time normalized by $T_{vol}$ for the HIT DENS data over two ensembles. The first is the full series of velocities over all trajectories. The second ensemble was obtained by truncating the velocity series of each trajectory such that only values measured within a certain volume of size $L$ were taken to mimic the finite volume effect. The figure shows that for times in the range $\tau < T_{vol}$, the difference that exists between the two autocorrelation estimations is rather small - up to $\sim 5\%$, meaning that the estimation of integral timescales within this range is reasonably close to that using the full range. For this reason, to estimate the Lagrangian integral timescales from the canopy trajectories, we used only time $\tau < 0.5 T_{vol}$, with $L = 3\si{\milli\meter}$ and $u'$ determined from all samples at a given sub-volume.

\section*{Appendix D - Estimation of 3d Eulerian Statistics}

\begin{figure}[H]
	\centering
	\includegraphics[width=7cm]{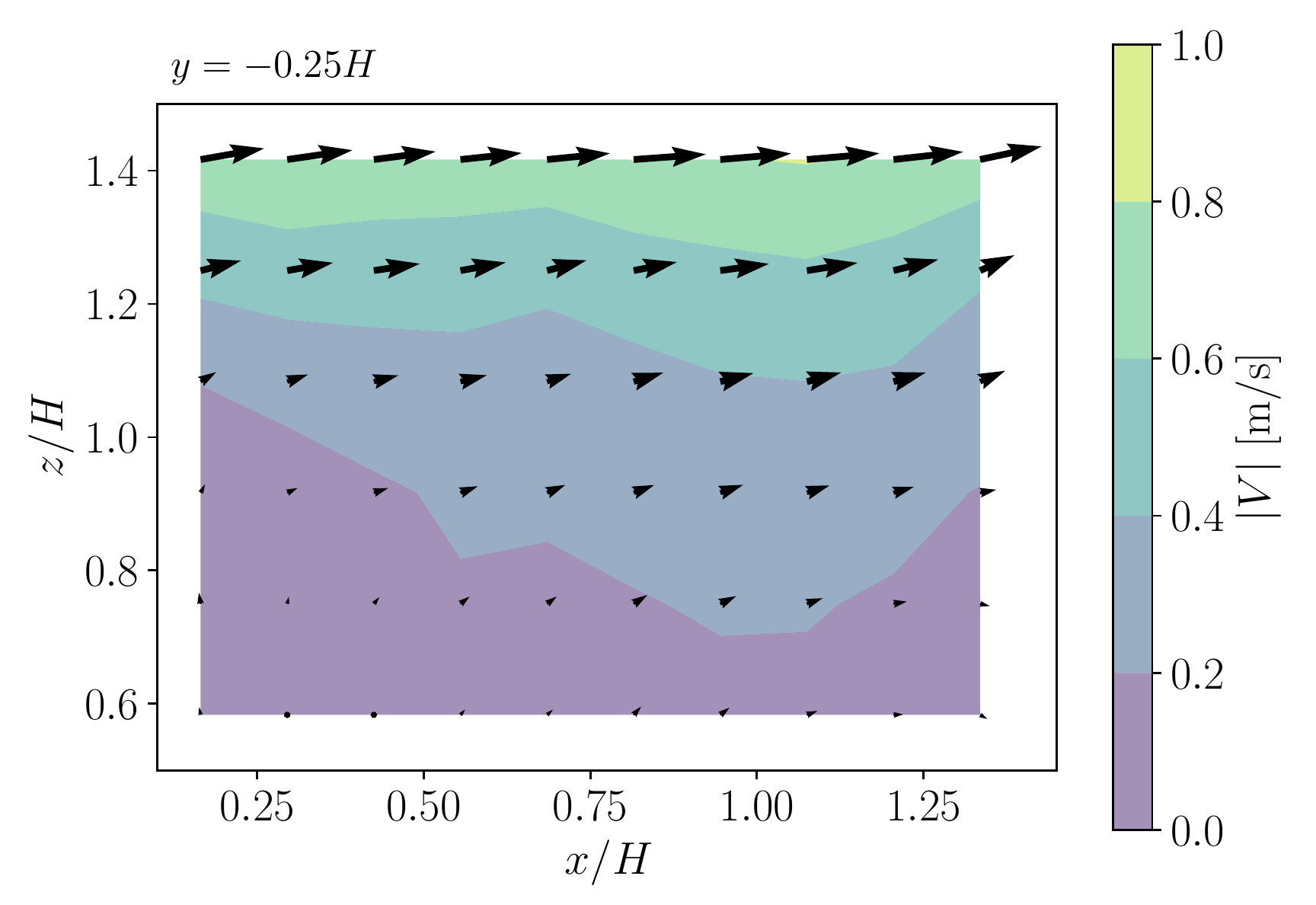}
	\caption{Two dimensional projection of the mean velocity over a plane parallel to the $(x, z)$ plane. The plane in found between two high roughness obstacles, intersecting sub-volumes a and b.  \label{fig:interp}}
\end{figure}

Using our Lagrangian dataset, we estimated Eulerian velocity statistics, such as the mean velocity and the turbulent stresses by three-dimensional interpolations of the sub-volume averaged data. The scheme was performed as follows: 
\begin{enumerate}
    \item we estimated a sub-volume averaged value for each desired statistics (e.g. $\av{u_i} = \av{v_i}$, and $R_{ij} = \av{ v'_i \, v'_j }$).
    \item we obtained vertical profiles of each statistics above each sub-volume group (a, b, c, d, see Fig.~\ref{fig:locs}(a)) by using a linear interpolation with respect to $z$.
    \item we obtain 3D field estimates by using a two-dimensional inverse distance weighted interpolation of the four vertical profiles.
\end{enumerate}
As an example, we present a two dimensional cut of the mean velocity field in Fig.~\ref{fig:interp}.

\end{document}